\documentclass[prd]{revtex4}

\usepackage{color,eucal,amsmath,amsfonts,amssymb,graphicx}

\def\bea{\begin{eqnarray}}
\def\beann{\begin{eqnarray*}}
\def\beq{\begin{equation}}
\def\eea{\end{eqnarray}}
\def\eeann{\end{eqnarray*}}
\def\eeq{\end{equation}}

\begin{document}

\title{Regular and chaotic interactions of two\ BPS dyons at low energy}
\author{Ricardo Fariello, Hilmar Forkel and Gast\~{a}o Krein}
\affiliation{Instituto de F\'{\i}sica Te\'{o}rica, Universidade Estadual Paulista, Rua
Pamplona, 145, 01405-900 S\~{a}o Paulo, SP, Brazil}

\begin{abstract}
We identify and analyze quasiperiodic and chaotic motion patterns in the
time evolution of a classical, non-Abelian Bogomol'nyi-Prasad-Sommerfield
(BPS) dyon pair at low energies. This system is amenable to the geodesic
approximation which restricts the underlying SU($2$) Yang-Mills-Higgs
dynamics to an eight-dimensional phase space. We numerically calculate a
representative set of long-time solutions to the corresponding Hamilton
equations and analyze quasiperiodic and chaotic phase space regions by means
of Poincar\'{e} surfaces of section, high-resolution power spectra and
Lyapunov exponents. Our results provide clear evidence for both
quasiperiodic and chaotic behavior and characterize it quantitatively.
Indications for intermittency are also discussed.
\end{abstract}

\author{}
\maketitle

\section{Introduction}

The classical dynamics of non-Abelian gauge theories is known to be chaotic
in a large part of its phase space \cite{bir94}. By itself this is not
unexpected since chaos is far more the rule than the exception in nonlinear
dynamical systems. Perhaps more surprising, however, is the mounting
evidence for this chaotic behavior, which is strictly speaking a classical
phenomenon, to be of relevance for physical quantum gauge theories as well.

Part of the existing indications for the chaoticity of non-Abelian gauge
theories stem from ``homogeneous approximations'' which neglect all spatial
variations of the fields. Although these drastic reductions of the dynamics
access only a tiny and not generally physical\ fraction of the full phase
space, the few remaining degrees of freedom proved sufficient to
establish chaotic regimes first in SU$\left( 2\right) $ Yang-Mills theory %
\cite{bas79} and later in Yang-Mills-Higgs theory \cite{mat81} and
Chern-Simons gauge theory \cite{gia88}. Extensive lattice calculations of
the time evolution under the full hyperbolic Yang-Mills equations
subsequently showed that spatially inhomogeneous gauge fields not only
evolve chaotically, too, but reveal even more complex and qualitatively
new phenomena \footnote{%
The chaoticity of gauge field trajectories was characterized by positive
maximal Lyapunov exponents \cite{mue92}, the whole Lyapunov spectrum \cite%
{gon94}, and the Kolmogorov-Sinai entropy \cite{bol00}. In combination,
these properties indicate the global hyperbolicity of classical non-Abelian
lattice gauge theory (in 3+1 dimensions). Other important findings were a
continuous cascading of the dynamical degrees of freedom (and their energy)
towards the ultraviolet during time evolution \cite{mue92}, which might
impede the continuum limit of non-Abelian lattice gauge theories \cite{nie96}%
, and the distribution of nearest-neighbor level spacings of Yang-Mills and
QCD lattice Dirac spectra according to Gaussian matrix ensembles \cite{hal95}%
, indicating from another perspective that the underlying classical theory
is chaotic \cite{boh84}.} which are of direct physical relevance, for
example, in nonequilibrium processes \footnote{%
At high temperatures, long-wavelength fields behave increasingly classical.
Chaos investigations can therefore provide useful input to the study of
otherwise hardly accessible nonequilibrium processes in hot gauge theories.
Timely applications include calculations of the fast thermalization rates
observed in heavy-ion collisions \cite{hei97} and (especially
topological)\ structure formation, e.g., in baryon
number violating processes during semi-classical evolution phases in the
standard model.}.

Between the above two computational extremes, i.e. either \emph{ad hoc} truncations
to only constant fields or the full solution of the classical field
equations on the lattice, there exist physically interesting subsystems of
the gauge dynamics whose spatially varying fields---typically soliton
configurations---and time evolution can be studied without invoking
uncontrolled approximations or requiring the solution of partial
differential equations. In the following, we will focus on a prototypical
such system, consisting of two electrically charged magnetic
Bogomol'nyi-Prasad-Sommerfield (BPS) monopoles \cite{pra75,bog76,sut97},
i.e. dyons, whose nontrivial spatial extension gives rise to crucial
properties including the magnetic charge. Nevertheless, the low-energy time
evolution of the dyon pair is accurately described by the geodesic
approximation which involves just a few collective degrees of freedom,
governed by ordinary differential equations. This geodesic dynamics is one
of the best understood examples for classical (and quantum) interactions
between extended\ solutions of physically interesting 3+1 dimensional field
theories and can be formulated at a rare level of explicitness \cite%
{GM:86,AH:88}. Beyond being fascinating in its own right (exhibiting e.g.
the celebrated scattering angle $\pi/2$ for head-on collisions), it
therefore serves as a paradigm for the interactions among many other
physically important solitons \footnote{%
Prominent examples include the interactions of Skyrmions, vortices,
instantons and D-branes.}. The purpose of the present paper is to examine
regular and chaotic motion patterns of this system in a representative set
of phase space regions.

Besides appearing in the electroweak sector of standard model extensions,
monopoles and their potentially chaotic interactions may have a crucial
function in the context of quark confinement by the strong
interactions. This becomes most explicit in the central role which BPS
monopoles play in the confinement mechanism of $N=2$ supersymmetric
Yang-Mills theory in 3+1 dimensions \cite{sei94}. By condensing in the
vacuum, they realize the classic 't Hooft-Mandelstam dual superconductivity
scenario \cite{tho76} in which color magnetic charges get screened
while color electric charges are confined by the dual
Mei\ss ner effect. Similar scenarios, in which the condensation of
monopole-like objects plays a key role, are expected to unfold in more
physical gauge theories as well \footnote{%
As a case in point, in the 2+1 dimensional Yang-Mills-Higgs model 't
Hooft-Polyakov monopoles (a generalization of BPS\ monopoles which play the
role of instantons in this case) generate ``weak confinement'' by forming a\
monopole antimonopole plasma, as shown long ago by Polyakov \cite{pol77},
and the ``strong confinement'' of 2+1 dimensional Yang-Mills theory is
expected to be due to a similar mechanism.}. In 3+1 dimensional Yang-Mills
theories, for example, there is lattice evidence for the condensation of Abelian-projected
monopoles to generate the bulk of the string tension \cite{che97}.
(According to an interesting recent suggestion, the ``active'' monopoles
might actually be BPS dyon constituents of caloron solutions with nontrivial
holonomy \cite{dia05}.) Hence monopoles may be instrumental in resolving the
most profound remaining mystery of QCD, i.e. the quark confinement mechanism %
\cite{mil}.

From a seemingly different but actually related perspective, quark
confinement is expected to be linked to the chaotic behavior of classical
chromodynamics as well. In fact, it has long been conjectured that the
vacuum of non-Abelian gauge theories, when undergoing a transition from
weakly to strongly coupled fields, also undergoes an order-disorder
transition and that the strongly coupled QCD vacuum is populated by highly
irregular color field configurations \cite{bir94}. In the limit of a large
number of colors, in particular, a vacuum made of random Yang-Mills fields
has been shown to be a necessary and sufficient condition for quark
confinement \cite{ole82}. From the outset, one of the motivations for
investigating chaos in non-Abelian gauge theories was therefore to shed
light on its potential role in the confinement mechanism \cite{bas79}.
Moreover, the instability of constant color-magnetic vacuum fields \cite%
{sav77} made it natural to conjecture that both gauge invariance and
stability of the physical vacuum may be restored by disordering the
color-magnetic background fields. Under the gluonic structures envisioned to
carry the bulk of this disorder are random domains as well as populations of
randomly distributed center vortices or monopoles.

The last of these scenarios may be related to the subject of our
investigation. Indeed, it is tempting to speculate that a disordered
ensemble of monopoles (and antimonopoles)\ in a semiclassical vacuum may be
generated by chaotic low-energy interactions among the monopoles. In the
following we are going to investigate precisely this type of interaction in
the simplest possible setting, i.e. between just two BPS monopoles. One may
then hope that the quantitative understanding of its chaotic regime will
lead to new insights into the disorder of interacting monopole ensembles as
well. For sufficiently dilute systems, expansions in the monopole density
and more sophisticated many-body techniques might even provide a starting
point for the quantitative treatment of chaotic multimonopole ensembles.
Alternatively, one could contemplate the technically challenging extension
of the geodesic approximation to approximate BPS multimonopole-multiantimonopole
solutions which incorporate multimonopole interactions \footnote{%
Such solutions may be similar or related to the BPS dyon constituents of
the recently discovered SU$\left( \text{N}\right) $ Yang-Mills caloron
solutions with nontrivial holonomy \cite{lee98}. (In the case of SU$\left(
2\right) $, they contain a BPS monopole-antimonopole pair.)}.

Based on the above motivations, our main objective in the following will be
to deepen the qualitative and quantitative understanding of quasiperiodicity
and chaos in the geodesic motion of two BPS dyons. The pioneering numerical
studies of this motion in Refs.~\cite{TR:88a,TR:89,TRA:93} already provided
several indications for its nonintegrability. (Evidence for chaotic
fluctuations around single monopoles in SU$\left( 2\right) $
Yang-Mills-Higgs theory exists as well \cite{joy92}, although not too large,
minimally spherically symmetric excitations remain regular \cite{for04}.) After a brief
recollection of the classical two-dyon dynamics at low energies, we will
first extend previous studies by examining Poincar\'{e} sections for a set
of numerically generated long-time trajectories of the dyon pair. In the
subsequent sections we break new ground by invoking high-resolution power
spectra and maximal Lyapunov exponents to analyze the dyon orbits further.
This analysis will go beyond the mere identification of standard motion
patterns and provide the first quantitative characterizations of
quasiperiodic and chaotic two-dyon trajectories.

\section{Classical low-energy dynamics of the two-dyon system}

\label{clasmo}

In order to set the stage for our investigation, we first recapitulate some
pertinent aspects of the dynamics of a classical BPS dyon pair at low
energies and discuss its known constants of the motion. Readers interested
in more detail are referred to the lucid discussion by Gibbons and Manton %
\cite{GM:86}.

We consider Yang-Mills-Higgs (YMH) theory in 3+1 dimensions with gauge group
SU($2$) and the Higgs field in the adjoint representation, i.e. the
Georgi-Glashow model. The classical field equations admit topological
soliton solutions which are magnetic monopoles with integer magnetic charge $%
k$ \cite{tho74}. In the following we will be interested in these monopole
solutions in the BPS limit of vanishing Higgs potential \cite{pra75,bog76,sut97},
which solve the more restrictive Bogomol'nyi equation \cite{bog76} 
\begin{equation}
B_{i}^{a}=\frac{1}{2}\varepsilon _{ijk}F_{jk}^{a}=\pm \left( \delta
^{ac}\partial _{i}+g\varepsilon ^{abc}A_{i}^{b}\right) \Phi ^{c}  \label{beq}
\end{equation}%
(where $F_{\mu \nu }^{a}$ is the field strength tensor of the gauge field $%
A_{\mu }^{a}$ and $\Phi ^{a}$ is the (adjoint) Higgs field). The solutions
of Eq.~(\ref{beq}) are the absolute minima of the YMH energy in their
topological charge sector and form a submanifold $M_{k}$ in
the space of gauge-inequivalent finite-energy fields. Static multimonopole
solutions (with $\left| k\right| >1$) of Eq.~(\ref{beq}) are possible
because the repulsive magnetic forces between the individual monopoles are
counterbalanced by the attractive forces which the massless Higgs field
mediates \cite{mon77}.

Although the underlying YMH dynamics is not directly embedded in the
standard model, it appears naturally in grand-unified theories. Monopole
solutions similar to the BPS prototype might therefore be physical. Their
mass would probably be large enough to explain why they have so far escaped
discovery in earthbound laboratories.

The monopole solutions come in families whose members are
characterized by continuous collective coordinates or ``moduli'' $x^{\alpha
} $. (For the one-monopole solution, for example, these are the three
position coordinates of the center and an overall phase angle.) Hence the moduli
space spanned by the collective coordinates\ is just the
(generally curved) manifold $M_{k}$. Its metric $g_{\alpha \beta }\!\left(
x\right) $ is induced by the metric on the more comprehensive space of all
finite-energy field configurations which the kinetic terms in the YMH
Lagrangian define.

The time evolution of a nonstatic and therefore in general electrically
charged BPS monopole system is governed by the hyperbolic partial
differential YMH equations whose quantitative analysis and
solution poses rather formidable problems. Reassuringly, however, the low-energy dynamics of BPS
dyons reduces to a much more tractable problem as long as Manton's
geodesic approximation \cite{NM:82} can be invoked. The latter rests on the
observation that energy conservation forces the low-energy motion of the $k$%
-dyon system (with sufficiently small initial velocities of the collective
coordinates, tangent to $M_{k}$) to remain close to one of the static BPS
solutions on $M_{k}$ at all times. This is simply because $M_{k}$ contains
all absolute YMH energy minima with magnetic charge $k$, so that moving out
of it would cost both kinetic and potential energy. Hence the low-energy
motion of a $k$-dyon system approximately corresponds to low-energy motion
of an associated point on the moduli space $M_{k}$. Since the energy of all
static $k$-monopole solutions is degenerate, i.e. at the same (minimal)
potential, furthermore, the low-energy motion is approximately determined by
the kinetic energy alone. Hence it corresponds to \emph{geodesic} motion on
the moduli space $M_{k}$, which is governed by the purely kinetic Lagrangian%
\begin{equation}
L_{geod}=\frac{m}{2}g_{\alpha \beta }\!\left( x\right) \dot{x}^{\alpha }\dot{x}%
^{\beta },  \label{geom}
\end{equation}%
where $m$ is the reduced mass of the dyons. Physically, this just means that
at small velocities (compared to the velocity of light) internal excitations
(vibrations) and deexcitations (radiation) can be neglected,
i.e. the dyons adapt adiabatically to their interactions by deforming
reversibly and scattering elastically. (Corrections to the geodesic
approximation were analyzed in an effective field theory framework in Ref.~%
\cite{bak98}.)

In the following, we will focus on the two-dyon system. Because of the
product structure of the moduli space, its center of mass momentum and an
overall phase (whose time dependence is associated with the total electric
charge) are individually conserved, and their metric is flat. Hence those
degrees of freedom decouple from the internal motion and can be separated
out. The remaining dynamics simplifies to the geodesic motion in the
four-dimensional internal part $M_{2}^{\left( 0\right) }$ of the moduli
space and can be studied independently. A physically intuitive coordinate
system on $M_{2}^{\left( 0\right) }$ consists of the Euler angles $\vartheta 
$, $\varphi $ and $\psi $, which determine the orientation of the two-dyon
system \footnote{%
At fixed time, the polar angles $\vartheta $ and $\varphi $ specify the
direction of the axis connecting the two monopoles, and $\psi $ corresponds
to rotations around this axis. The angle $\psi $ becomes relevant since the BPS
monopole pair is in general not axially symmetric.}, and the distance
variable $\varrho $ which measures the separation between the two dyon
centers (at large $\varrho $).

Although the metric on $M_{2}^{\left( 0\right) }$ is induced by the SU$%
\left( 2\right) $ YMH dynamics, its direct derivation from
the kinetic terms of the YMH Lagrangian seems out of reach. Instead, it has been
constructed on the basis of ingenious symmetry arguments by
Atiyah and Hitchin (AH) \cite{AH:85}, as summarized in Appendix \ref{ahm}.
Specialization of Eq.~(\ref{geom}) to the AH metric then determines
the geodesic dynamics of the dyon pair explicitly. The resulting Lagrangian%
\begin{equation}
L_{AH}=\frac{1}{2}\left[ f^{2}\!\left( \varrho \right) \dot{\varrho}%
^{2}+a^{2}\!\left( \varrho \right) \omega _{x}^{2}+b^{2}\!\left( \varrho \right)
\omega _{y}^{2}+c^{2}\!\left( \varrho \right) \omega _{z}^{2}\right]
\label{lah}
\end{equation}%
turns out to be nonlinear since the AH metric is curved. The functions $a$, $%
b$, $c$ and $f$ of the separation $\varrho $ are given in Appendix \ref{ahm} and
the $\omega _{i}\!\left( \vartheta ,\varphi ,\psi \right) $ are angular
velocities of the monopole (or dyon) pair around the axes of the body-fixed
frame,%
\begin{align}
\omega _{x}& =-\dot{\vartheta}\sin {\psi }\,+\dot{\varphi}\sin {\vartheta }%
\cos {\psi }\,, \\
\omega _{y}& =\dot{\vartheta}\cos {\psi }\,+\dot{\varphi}\sin {\vartheta }%
\sin {\psi }\,, \\
\omega _{z}& =\dot{\psi}+\dot{\varphi}\cos {\vartheta }\,.
\end{align}%
The form of the Lagrangian (\ref{lah}) is analogous to that of a nonrigid
body with distance-dependent ``moments of inertia'' $a^{2},b^{2}$ and $c^{2}$
around the body-fixed axes. Following Gibbons and Manton, we define the
radial coordinate $\varrho $ by choosing $f=-b/\varrho $ which leads to
convenient expressions for $a,b$ and $c$ \cite{GM:86}. In Appendix \ref%
{eiom}, the four Euler-Lagrange equations of the geodesic motion are derived
by variation of Eq.~(\ref{lah}).

For the question of integrability versus chaos, the number of integrals of
the motion plays a decisive role. In fact, a
motion is (Liouville) integrable only if the number of independent conserved
quantities (whose mutual Poisson brackets vanish) at least matches the
number of degrees of freedom. For the geodesic dynamics (\ref{lah}), three
constants of the motion are known explicitly (cf. Appendix \ref{ioms}), namely,
the total angular momentum%
\begin{equation}
\mathcal{M}^{2}=p_{\vartheta }^{2}-2p_{\varphi }p_{\psi }\cot {\vartheta }%
\csc {\vartheta }+\left( p_{\varphi }^{2}+p_{\psi }^{2}\right) \csc ^{2}{%
\!\vartheta }
\end{equation}%
(where the $p_{\alpha }$ are generalized momenta canonically conjugate to
the coordinates $\alpha $, as defined in Eq.~(\ref{eqn:ps}) of Appendix \ref%
{eiom}), the energy (\ref{h}) which, for geodesic motion, equals the Lagrangian since
the potential on $M_{2}^{\left( 0\right) }$ vanishes,%
\begin{equation}
E_{AH}=\frac{1}{2}\left( \frac{p_{\varrho }^{2}}{f^{2}}+a^{2}\omega
_{x}^{2}+b^{2}\omega _{y}^{2}+c^{2}\omega _{z}^{2}\right)  \label{eah}
\end{equation}%
and, finally, the generalized momentum $p_{\varphi }$
conjugate to the coordinate $\varphi $ which is cyclic, i.e. does not appear
explicitly in Eq.~(\ref{lah}).

At least one additional, independent constant of the motion is therefore
required for the two-dyon motion to become integrable. Such a fourth
conserved quantity indeed exists (to arbitrarily good approximation) in at
least one region of phase space, namely, where the two dyons remain
infinitely separated. Since they cannot exchange electric charge then 
\footnote{%
despite the long-range forces generated by the massless Higgs fields}, the time
evolution of each of their phases\ or, equivalently, their \emph{individual}
electric charge is conserved. Hence the motion of two far separated dyons must be
integrable, as intuitively expected, and cannot exhibit chaos. (At
asymptotic distances the AH metric reduces to the Euclidean Taub-NUT metric
whose geodesic motion is indeed known to be integrable \cite{GM:86,vam96}.)

This situation changes, however, if the two dyons begin to approach each
other. Only their total charge, but not their charge ratio, remains
conserved since the dyons are then able to exchange charge through the
Higgs field. (Even when starting asymptotically with two uncharged
monopoles, one will therefore generally end up with two dyons of opposite
but nonvanishing charge.) If the integral of the motion associated with the
relative phase ceases to exist, chaotic motion becomes possible \footnote{%
Recall that this motion is not ergodic since the remaining three constants
of motion prevent it from filling the seven-dimensional constant-energy
surface densely and uniformely.}. It is this region of the phase space in
which we will be particularly interested below.

The above observations raise the question for which orbits the
violation of individual charge conservation can typically be considered as a
small perturbation away from the integrable Taub-NUT limit. In our context,
this question is of relevance because for such orbits the transition from
integrable to\ chaotic motion will be delayed and potentially obscured by
the implications of KAM theory. Indeed, the KAM theorem \cite{kam} states
that almost all invariant tori of the unperturbed Taub-NUT motion will
remain intact during sufficiently small deviations from the $\varrho
\rightarrow \infty $ limit \footnote{%
The existence of bounded quasiperiodic geodesics in the almost asymptotic\
AH metric, i.e. the existence of quasiperiodic solutions to the dynamics (%
\ref{lah}) in its sufficiently weakly nonintegrable realms, has been
established in Ref.~\cite{MW:88} and shown to form a set with positive
Lebesgue measure.}. Under weakly nonintegrable perturbations the motion
should therefore stay quasiperiodic for almost all initial conditions.
Chaotic motion will then be restricted to the small set of trajectories which lie
on the descendants of invariant tori with commensurate frequencies.

Although we have argued above that it is unlikely for the geodesic AH motion
to be integrable outside of the asymptotic region, the existence of
additional but so far undiscovered constants of the motion cannot be
excluded \emph{a priori} since no general method for finding \emph{all}
conserved quantities of a nonlinear dynamical system is available. Hence it
was an important step by Temple-Raston and Alexander to gather the first
numerical evidence for the existence of chaotic regions in the two-dyon
phase space \cite{TR:88a,TR:89,TRA:93}. In the next section, we will
elaborate on part of these results by extending the Poincar\'{e} section
analysis of Ref.~\cite{TR:89} and by looking for orbits
which bear the insignia of chaos.

\section{Poincar\'{e} sections}

\label{psec}

We begin our search for chaotic regions in the phase space of the dyon pair
by analyzing the Poincar\'{e} sections of several typical
trajectories.\ A\ Poincar\'{e} section draws a selective portrait of a given
phase space orbit (over a finite time interval) which achieves an enormous
data reduction by mapping the whole trajectory into a discrete set of
points. This set transparently exhibits characteristic global aspects of the
orbit and allows, in particular, a direct visual distinction between orbits
which arise from integrable (or weakly nonintegrable, in the KAM sense) and
fully nonintegrable dynamics.

Poincar\'{e} section analyses are most powerful for conservative systems
with two degrees of freedom where a transparent graphical interpretation
becomes possible. At first glance one might therefore doubt their utility in
the eight-dimensional phase space of the two-dyon problem. However, in Ref.~%
\cite{TR:89} it was recognized that a particular canonical transformation
turns a second constant of the dyon pair motion (besides the generalized
momentum $p_{\varphi }$), namely, the total angular momentum squared $%
\mathcal{M}^{2}$, into a canonical momentum. As a consequence, the
transformed Hamiltonian becomes independent of the associated canonical
variables while $p_{\varphi }$ and $\mathcal{M}^{2}$ act as fixed
``external'' parameters. This Hamiltonian actually defines a reduced,
four-dimensional (and still symplectic) phase space in which $p_{\varphi }$
and $\mathcal{M}^{2}$ are automatically conserved at every point. Energy
conservation further constrains all orbits to a three-dimensional
hypersurface inside this reduced phase space and thus makes a graphical
Poincar\'{e} section analysis feasible.

The Poincar\'{e} section of any given orbit is the set of its intersection
points, in a fixed direction, with a suitable two-dimensional plane in phase
space. In our case, a useful choice \cite{TR:89} is the $\left( \varrho
,p_{\varrho }\right) $ plane located at the position $\mathcal{M}_{1}=0$
which we will adopt throughout. Hence we define the elements of an orbit's
Poincar\'{e} section as those points in which it pierces through this plane
while $\mathcal{M}_{1}$ changes from positive to negative sign.

One can easily convince oneself that integrable and chaotic motions generate
qualitatively different Poincar\'{e} sections. Indeed, the existence of an
additional, fourth constant of the motion $C_{4}$ (as provided e.g. by $%
p_{\psi }$ for far separated dyons, cf. Appendix \ref{ioms}) would restrict all
orbits to a generally two-dimensional submanifold in phase space, namely, the
intersection of the two three-dimensional hypersurfaces on which either the
energy or $C_{4}$ have fixed values. The dyon-pair motion then becomes
(Liouville) integrable, the submanifold becomes an invariant torus and the
orbit's intersection points with the $\left( \varrho ,p_{\varrho }\right) $
plane sweep out a (maximally) one-dimensional curve. If the system is
nonintegrable, on the other hand, no fourth constant of the motion exists
and the orbit is only bound to the three-dimensional constant-energy surface
(for strong enough nonintegrability in the KAM\ sense). Its intersection
points with the $\left( \varrho ,p_{\varrho }\right) $ plane can\ therefore
be more broadly distributed and eventually fill a two-dimensional area,
which provides a clear signature for a chaotic dynamical regime \footnote{%
More specifically, a simple periodic motion leads to a single fixed point in
the Poincar\'{e} section while a periodic orbit with two commensurable
frequencies (whose quotient is a rational number) gives rise to a finite
number of points being indefinitely repeated in the same order. A
two-frequency quasiperiodic orbit draws a closed curve that never exactly
retraces its steps, and chaotic motion appear as a scatter of points which
eventually fill a two-dimensional surface.}.

In order to prepare for the Poincar\'{e} section analysis, we have generated
twelve representative orbits by numerically integrating the equations of
motion (\ref{eqn:movf1})--(\ref{eqn:movf5}) under suitable initial
conditions \footnote{%
In the large-$\varrho $ region, the dyon-pair orbits can be obtained
analytically by perturbation of the asymptotic Taub-NUT metric and turn out
to be conic sections \cite{GM:86}. It would therefore be possible to obtain
analytical expressions for the Poincar\'{e} sections in this region, too.
Since we expect the chaotic regime to develop at smaller $\varrho $ values,
i.e. outside the range of validity of the (first-order)\ analytical results,
however, we will resort to a numerical treatment of all orbits.}. The latter
are provided by specifying initial values for the four coordinates and their
time derivatives. In order to facilitate the comparison with the results of
Ref.~\cite{TR:89}, we adopt the following five initial conditions, and the
form of a sixth, at the initial time $t_{0}$: 
\begin{equation}
\vartheta =\frac{\pi }{2},\hspace{0.5cm}\varphi =\psi =0,\hspace{0.5cm}\dot{%
\varrho}=\dot{\vartheta}=0\hspace{0.5cm}\text{and}\hspace{0.5cm}\dot{\varphi}%
=\frac{h_{2}}{a^{2}}.  \label{icond1}
\end{equation}%
As a consequence of Eqs.~\eqref{eqn:ps}, the corresponding initial values of
the conjugate momenta are 
\begin{equation}
p_{\varrho }=0,\hspace{0.5cm}p_{\vartheta }=\mathcal{M}_{2}=0,\hspace{0.5cm}%
p_{\varphi }=\mathcal{M}_{1}=h_{2}\hspace{0.5cm}\text{and}\hspace{0.5cm}%
p_{\psi }=\mathcal{M}_{3}=c^{2}\dot{\psi},
\end{equation}%
which implies 
\begin{equation}
\mathcal{M}^{2}\equiv \mathcal{M}_{1}^{2}+\mathcal{M}_{2}^{2}+\mathcal{M}%
_{3}^{2}=p_{\varphi }^{2}+p_{\vartheta }^{2}+p_{\psi }^{2}=h_{2}^{2}+c^{4}\dot{\psi}^{2}.  \label{m2bis}
\end{equation}%
Here $h_{2}$ is the parameter in the reduced Hamiltonian which fixes the
value of the conserved momentum $p_{\varphi }$. Hence only the initial
values for $\dot{\psi}$ and $\varrho $ remain to be determined. If $h_{2}$
is given, Eq.~(\ref{m2bis}) implies that the initial condition for $\dot{\psi%
}$ can be specified by fixing the value of another conserved quantity,
namely, the total angular momentum squared $\mathcal{M}^{2}$. Finally,
instead of prescribing the initial condition for $\varrho $ directly, it is
more convenient to use the reparametrization%
\begin{equation}
\varrho =2K\!\left( \sin {\!\frac{\beta }{2}}\right)
\end{equation}%
($K$ is an elliptic integral, cf. Eq.~(\ref{ke})), which maps infinite dyon
separation ($\varrho \rightarrow \infty $) into $\beta =\pi $, and to
specify the initial value of $\beta $. Under the above conditions (\ref%
{icond1}), each orbit is therefore uniquely characterized by the values of $%
\beta $ and two integrals of the motion, $h_{2}$ and $\mathcal{M}^{2}$. For
the interpretation of the results below it will be helpful to keep in mind that
(at fixed $h_{2}$) the \emph{initial} relative electric charge $p_{\psi
}\!\left( t_{0}\right) $ of the two dyons, and therefore their initial Coulomb
interaction, grows with $\mathcal{M}^{2}$ (cf. Eq.~(\ref{m2bis})).

A fourth-order Runge-Kutta algorithm~\cite{GG:89} was used for the numerical
integration of the equations of motion. The Eqs.~\eqref{eqn:abc} are
solved along the way to obtain the values of the AH functions $a$, $b$ and $%
c $ at each $\varrho $. The length of an integration step was typically $10^{-4}$ and the
longest runs contained $2^{27}$ steps. The accuracy of the generated orbits
was monitored after each time step by calculating the three conserved
quantities (i.e. $E$, $\mathcal{M}^{2}$ and $p_{\varphi }$). The maximal
deviations from their fixed values were of the order $10^{-12}$. The Poincar%
\'{e} section of a given orbit is then constructed from those of its points at which
$\mathcal{M}_{1}$ vanishes and $\mathcal{\dot{M}}_{1}<0$ \footnote{%
Note that the times between consecutive piercings of the $\mathcal{M}_{1}=0$
plane by an orbit typically vary. This is in contrast to stroboscopic
studies where one determines the system's output at equal time intervals.}.
Since the numerical integration routine generates the orbits only at
discrete time steps, the exact intersection points were determined by\
polynomial (Lagrange) interpolation between the two adjacent points on the
orbit with opposite signs of $\mathcal{M}_{1}$.

We will now discuss the Poincar\'{e} sections for twelve orbits
which all share the same $h_{2}$ value, $h_{2}=37.596$. We arrange them into
three plots, each containing the sections of four orbits with a common value
of $\mathcal{M}^{2}$ and initial values $\beta =\left(
3.11,3.12,3.13,3.14\right) $ of the radial coordinate. The corresponding
initial values for $\left( \varrho ,p_{\varrho }\right) $ are indicated by a
small circle although $\mathcal{M}_{1}=h_{2}\neq 0$
implies that they do not lie in the $\mathcal{M}_{1}=0$ plane. Each of the
twelve selected orbits describes charged\ monopole, i.e. dyon interactions
since $\mathcal{M}^{2}>h_{2}^{2}=\allowbreak 1413.5$ and Eq.~(\ref{m2bis})
imply that the initial value of their momenta $p_{\psi }=c^{2}\dot{\psi}$,
associated with the relative electric charge, is nonzero. The above initial
conditions were chosen to yield a rather representative set of Poincar\'{e}
sections similar to those considered in Ref.~\cite{TR:89}, in order to both
qualitatively confirm the results obtained there and to extend them into
neighboring phase space regions.

The first plot, Fig.~\ref{fig:pp01}, contains the four Poincar\'{e} sections
with the smallest value $\mathcal{M}^{2}=1906.71$, i.e. with the weakest
initial Coulomb attraction between the dyons. This explains why each of the
orbits covers a rather large range of $\varrho $ values while the variation
of the radial velocities, i.e. the $p_{\varrho }$ range, is
relatively moderate. In three of the sections, furthermore, the two dyons
never come closer than their initial distance, and in all four their
separation $\varrho \gtrsim 11$ seems to stay well inside the asymptotic
region where the relative charge of the dyons becomes (almost)\ time-independent
and the motion (approximately or KAM-)\ integrable. This is
confirmed by the Poincar\'{e} sections whose points indeed trace a
one-dimensional closed curve which, within plot resolution, appears
mostly continuous. Poincar\'{e} sections of this type are generated by
superpositions of periodic motions with incommensurate frequencies, i.e. by
quasiperiodic orbits. As discussed above, quasiperiodic behavior indicates
that the dynamics is either integrable or at most weakly nonintegrable in
the KAM sense.

In the next plot, Fig.~\ref{fig:pp02}, we display the analogous Poincar\'{e}
sections with a larger value of $\mathcal{M}^{2}=2152.95$ and therefore with
a stronger initial Coulomb attraction. Comparison with the sections of Fig.~%
\ref{fig:pp01} shows that the variations in dyon distance are now smaller
(tighter orbits) while their relative momenta vary more strongly over each
of the orbits. The increased attraction also brings the dyons closer
together, their maximal separation now being the initial one for all four
orbits. Nevertheless, their minimal separation $\varrho _{\min }\sim 8$
seems to stay large enough for the motions to remain approximately
integrable since all four Poincar\'{e} sections still form one-dimensional,
closed curves.

We therefore increase $\mathcal{M}^{2}$ further, to $\mathcal{M}^{2}=2237.70$%
, and plot the corresponding Poincar\'{e} sections in Fig.~\ref{fig:pp03}.
Clearly, the character of the outermost section differs qualitatively from
all those encountered previously. Instead of remaining constrained to a
curve, it visibly spreads out into a two-dimensional area of the $\left(
\varrho ,p_{\varrho }\right) $ plane. Hence it corresponds to an aperiodic
orbit and strongly suggests that the dynamics has become chaotic. This
interpretation is consistent with the fact that the two dyons approach each
other the most in this orbit. Their minimal distance $\varrho _{\min }\sim
2\pi $, which is about twice the distance of the bolt singularity at $%
\varrho =\pi $, apparently suffices for electric charge exchange by the
Higgs field to become efficient and hence for $p_{\psi }$ to cease being
even an approximate integral of the motion \footnote{%
Numerically, this can be checked by calculating the time dependence of $%
p_{\psi }$ along the trajectory. For the chaotic orbit it is indeed much stronger
then for all previous ones.}. This confirms earlier indications
for the chaoticity of the geodesic two-dyon dynamics \cite{TR:88a} which
were supported by the analysis of similar Poincar\'{e} sections %
\cite{TR:89}, Julia sets \cite{TR:88b} and escape plots of two-monopole
scattering trajectories \cite{TRA:93}. An additional chaotic orbit, with a
still larger value of $\mathcal{M}^{2}$, will be generated and analyzed in
the following sections.

\section{Power spectra}

\label{psp}

Numerical studies of a single chaos indicator are generally not sufficient
to establish chaotic behavior with certainty. Inevitable numerical roundoff
errors as well as specific features of a system under consideration (e.g.
particular regions of instability)\ make it often desirable to probe the
character of the motion by several complementary techniques. Typical pitfalls,
like the premature misinterpretation of a seemingly irregular but perfectly
integrable quasiperiodic motion pattern as chaos, can thereby be avoided.

Power spectral analysis has proven particularly useful for the distinction
between (quasi-) periodic and chaotic time evolution \cite%
{GS:75;CFPS:80;LPC:81;BJ:87}. This is because the power spectra of ordered
motion (either periodic or quasiperiodic) consist of sharp resonance lines
which appear at simple harmonics of the base frequency in the periodic case
and at any linear combination of all integer multiples of the base
frequencies in the quasiperiodic case. Aperiodic systems, in contrast, are
generally chaotic and have continuous and noisy power spectra.

In order to sharpen and extend the interpretation of the Poincar\'{e}
section analysis of Sec.~\ref{psec}, we have submitted the underlying
orbit data to a spectral analysis. These data are four-dimensional,
discrete time series of specific orbit solutions on $M_{2}^{\left( 0\right)
} $ whose coordinates $\left\{ x^{\alpha }\!\left( t_{k}\right) \right\}
=\left\{ \vartheta ,\varphi ,\psi ,\varrho \right\} $ are
recorded numerically at equally spaced times $t_{k}:=k\Delta t$. Based on
these solutions, one can calculate the time evolution of any dynamical
variable $f\!\left( \left\{ x^{\alpha }\!\left( t_{k}\right) \right\} \right) $
of interest as well as its discrete Fourier transform. We found it useful to
choose the generalized momentum $p_{\varrho }\!\left( t_{k}\right)
=f^{2}\!\left( \varrho \!\left( t_{k}\right) \right) \dot{\varrho}\!\left(
t_{k}\right) $ conjugate to $\varrho $ as the dynamical variable. It has the
power spectrum 
\begin{equation}
P_{p_{\varrho }}\!\left( \nu \right) =\left| \frac{1}{\sqrt{N}}%
\sum_{k=0}^{N-1}p_{\varrho }\!\left( k\Delta t\right) e^{-2\pi ik\nu
/N}\right| ^{2} \label{pwspec}
\end{equation}%
which is a function of the mode frequencies $\nu $. (The Wiener-Khinchin
theorem asserts that the alternative definition of $P\!\left( \nu \right) $ as
the Fourier transform of the time series' autocorrelation function is
equivalent if the correlations decay sufficiently fast.) In order to assess
a potential bias due to the choice of $p_{\varrho }$ as the dynamical
variable, we have also calculated the power spectra associated with the time
evolution of $\varrho $ and obtained essentially analogous results.

We numerically compute the power spectra by means of the Fast Fourier
Transform algorithm. Appropriate windowing is used to suppress artificial
oscillations due to the discrete data set and the finite time
interval (see Ref.~\cite{DK:00} and references therein). Since the impact of
these artifacts decreases with the length of the time interval over which
data for $p_{\varrho }\!\left( t_{k}\right) $ are available, we sample the
orbits after each of $2^{25}$ time steps of length $\Delta t=10^{-4}$ which
ensures sufficient accuracy for our purposes. We have calculated power
spectra for the\ three orbits associated with the outermost Poincar\'{e}
sections in each of the Figs.~\ref{fig:pp01}--\ref{fig:pp03}, specified by
the initial conditions given in Sec.~\ref{psec}, and for an additional orbit
to be discussed below.

The $p_{\varrho }\!\left( t_{k}\right) $ time series based on the orbit
associated with the outermost Poincar\'{e} section of Fig.~\ref{fig:pp01} is
depicted in Fig.~\ref{fig:qtps}a over $4 \times 10^{6}$ time steps. The
corresponding power spectrum (obtained from all data on $p_{\varrho}$, i.e.
from the full orbit with $2^{25}$ time steps over a total time of $3.35 \times 10^{3}$)
is shown in a relatively small but representative frequency interval
in Fig.~\ref{fig:qtps}e. On the basis of its Poincar\'{e} section, the
underlying orbit was interpreted as quasiperiodic in Sec.~\ref{psec}. The
power spectrum confirms this interpretation but contains much more
quantitative information. Indeed, while the underlying time series in Fig.~%
\ref{fig:qtps}a is in many ways indistinguishable from a periodic one, its
power spectrum exhibits sharp peaks \footnote{%
Note that we plot the logarithm of the power spectra in order to emphasize
the structure of the background.} of varying strength which are located at odd integer
linear combinations of two rationally independent (or incommensurate) 
\footnote{%
Rational independence of $\nu _{1,2}$ implies that the only solution of $%
m_{1}\nu _{1}+m_{2}\nu _{2}=0$ is $m_{1}=m_{2}=0$.} base frequencies $\nu
_{1}\simeq 0.0653$ and $\nu _{2}\simeq 0.0978$. Hence this orbit is two-mode
quasiperiodic \footnote{%
The choice of the base frequencies is to a certain extent a matter of
convention \cite{ER:85}.}.

We now turn to the orbit whose Poincar\'{e} section is the outermost in Fig.~%
\ref{fig:pp02}. Again, its time series (cf. Fig.~\ref{fig:qtps}b) is
visually difficult to distinguish from periodic motion while its power
spectrum in Fig.~\ref{fig:qtps}f reveals a quasiperiodic motion with the two
incommensurate base frequencies $\nu _{1}\simeq 0.1234$ and $\nu _{2}\simeq
0.1848$. In contrast to the previous power spectrum, however, the peaks at
frequencies corresponding to more complex linear combinations of $\nu _{1}$
and $\nu _{2}$ are practically undetectable here. This seems to be a
consequence of the larger initial coupling between the two dyons and is a
rather frequent occurrence among sufficiently strongly interacting
quasiperiodic systems. Although experimentally confirmed in many situations,
a rigorous theoretical explanation for this type of behavior appears still to be
missing \cite{ER:85}.

As foreshadowed by the results of the Poincar\'{e} section
analysis, a strikingly different power spectrum is obtained from the orbit
associated with the outermost section of Fig.~\ref{fig:pp03}. Indeed, this
is the orbit whose Poincar\'{e} section indicated the chaoticity of the
geodesic dynamics. Although the time dependence of its $p_{\varrho }$ in
Fig.~\ref{fig:qtps}c remains, apart from small and irregular modulations,
similar to the previous cases, the corresponding power spectrum in Fig.~\ref%
{fig:qtps}g has a qualitatively different character. Instead of sharp and
isolated peaks, it now contains equally spaced, broadened peaks on top of a
smooth and (within our frequency resolution) continuous background power
distribution \footnote{%
We recall the well-known effect that in a continuous power
spectrum based on a finite time interval the contributions from the lowest frequencies are artificially
enhanced. This is due to the fact that aperiodic points in a finite data set
appear as points with very long periods, comparable to the length of time
over which the series is recorded.}, covering all recorded frequencies and
signalling the onset of aperiodicity \footnote{%
In fact, it appears that the quasiregular orbits unfold the skeleton of the
aperiodic signals. A further analysis of such orbits could be persued by the
method of Ref.~\cite{SBP:89} which allows to separate the power spectra of
conservative systems into sharp and broadband features by utilizing the
distribution of local Lyapunov exponents \cite{GBP:88}.}.

The obvious departure from quasiperiodic behavior strengthens the evidence
for the chaoticity of this orbit. Our discussion in Sec.~\ref{clasmo}
indicates, furthermore, that even the quasiperiodic behavior of the previous
spectra, especially that of Fig.~\ref{fig:qtps}f, may have been generated
by orbits which lie outside of the strictly integrable phase space region
cor\-res\-pon\-ding to asymptotically far separated dyons. Indeed, a discrete
power spectrum still results if the nonintegrability is caused by weak
perturbations and if the orbit remains on a KAM torus, i.e. continues to
behave quasiperiodically \footnote{%
It would be interesting to study the onset of chaos further, e.g., by
numerically identifying orbits on commensurate-frequency KAM tori which
nonintegrable perturbations destroy first.}. Our results imply that such a
``delayed'' onset of chaos might persist down to minimal distances $\varrho
_{\min }\gtrsim 8$ between the two dyons.

In any case, the above findings suggest that the faster and closer\ the two
dyons approach, the more unpredictable, i.e. chaotic their motion becomes.
In order to test and extend this conclusion, we have calculated a further
orbit with a still larger total angular momentum squared (and consequently
stronger initial Coulomb interaction), $\mathcal{M}^{2}=2359.46$, and a
somewhat smaller initial distance $\beta =3.13$ between the dyons than in
the previous chaotic orbit. Its power spectrum, displayed in Fig.~\ref%
{fig:qtps}h, has a dense background and confirms the expectation that the
new orbit is aperiodic as well. Comparison with the power spectrum of the
first chaotic orbit shows, furthermore, that the background in Fig.~\ref%
{fig:qtps}h is less noisy and that more sharp frequencies remain clearly
discernible than in Fig.~\ref{fig:qtps}g. The faster time dependence of the
associated $p_{\varrho }\!\left( t\right) $ in Fig.~\ref{fig:qtps}d, on the
other hand, exhibits more pronounced amplitude modulations. In the following
section we will continue the analysis of the two chaotic orbits by
calculating their maximal Lyapunov exponents.

One might perhaps wonder why we have encountered no quasiperiodic behavior
with more than two fundamental modes. This is not particularly surprising,
however, since two-mode quasiperiodicity is the rule rather than the
exception among sufficiently strongly coupled, nonlinear dynamical systems.
Indeed, the nonlinear couplings between an increasing number of modes
tend to replace quasiperiodicity by chaos \cite{rue71}. It is
quite plausible that this happens in the two-dyon system as well and
explains the predominance of quasiperiodic orbits with the minimal number of
two base frequencies.

In summary, our spectral analysis has uncovered that those parts of the
two-dyon phase space which remain close enough to the asymptotic $\varrho
\rightarrow \infty $ region are characterized by two-frequency quasiperiodic
motion. In addition, we have determined the base frequencies of two
typical quasiperiodic orbits quantitatively. Perhaps most importantly, the power
spectra have also accomplished their main task to separate quasiperiodic
from irregular behavior. In particular, they strongly support the
identification of two orbits, the last one from Sec.~\ref{psec} and an
additional one with a still larger initial Coulomb force between the dyons,
as chaotic. Our findings therefore substantially increase previous evidence
that, apart from the asymptotic $\varrho \rightarrow \infty $ region, the
relative low-energy motion of two BPS dyons admits only three independent
conserved quantities and turns out to be genuinely nonintegrable.

While the distinction between (quasi-) periodic and chaotic motion is a
particular strength of the spectral analysis, it does relatively little to
further characterize chaotic behavior \footnote{%
This holds even more strongly for dissipative systems with strange attractors
because power spectra discard the phase information of the Fourier spectrum
(cf. Eq.~(\ref{pwspec})).}. Hence we consider it useful to subject our
orbits to yet another classic analysis tool from the arsenal of chaos
indicators, by calculating their characteristic or Lyapunov exponents. These
exponents complement our previous analyses particularly well since they are
specifically designed to quantify the chaoticity of irregular motion patterns.

\section{Lyapunov exponents}

\label{lce}

While the high-resolution spectral analysis of the last section has very
clearly distinguished quasiperiodic from irregular orbits and
determined the type of quasiperiodic behavior and its base frequencies quantitatively, it
is much less specific about the properties of the aperiodic motion. It
cannot, for example, unambiguously distinguish chaotic from (quasi-)\ random
behavior (potentially due to roundoff errors).

Although previous work and our results of the last two sections have
produced strong evidence for the chaoticity of two-dyon orbits in
specific regions of the (finite $\varrho $) phase space, it therefore
remains desirable to carry the investigation a step further and to obtain a
quantitative characterization of the chaotic behavior. The most fundamental
such characterization is provided by the values of positive Lyapunov
exponents \cite{BGS:76}. After a brief and informal introduction to the
underlying concepts, we will therefore evaluate the largest of
these characteristic exponents for selected dyon pair orbits.

The primary function of Lyapunov exponents is to quantify the logarithmic rate
of convergence or divergence between two orbits which started at an initial
time $t_{0}$ at neighboring positions ${\mathbf{x}}_{0}$ and ${\mathbf{x}}%
_{0}+\delta {\mathbf{x}}_{0}$ (with $\lVert \delta {\mathbf{x}}_{0}\rVert
\ll 1$ where $\lVert \cdot \rVert $ denotes the norm with respect to a
Riemannian metric). Their relevance is obvious: exponentially divergent
orbits, corresponding to positive Lyapunov exponents and signalling an
exponential sensitivity of the time evolution to the initial conditions, are
the prototypical signature of chaos. More specifically, under dynamical
evolution for a sufficiently long time $t$ the deviation between the
position of both orbits becomes $\delta {\mathbf{x}}_{t}$ and the maximal
Lyapunov exponent $\mathsf{L}_{\text{max}}$ is defined by relating the deviation norms
as $\lVert \delta {\mathbf{x}}%
_{t}\rVert \approx \exp \!\left( \mathsf{L}_{\text{max}}t\right) \lVert \delta
{\mathbf{x}}_{0}\rVert $, i.e.
\begin{equation}
\mathsf{L}_{\text{max}}=\lim_{t\rightarrow \infty }\lim_{\lVert \delta {%
\mathbf{x}}_{0}\rVert \rightarrow 0}\frac{1}{t}\ln \!\frac{\lVert \delta {%
\mathbf{x}}_{t}\rVert }{\lVert \delta {\mathbf{x}}_{0}\rVert }.
\label{eqn:lcemax-def}
\end{equation}%
For a detailed and rigorous treatment of Lyapunov exponents we refer to
original work by Oseledec \cite{VIO:68} and Ruelle \cite{DR:79}, as well as
to the review \cite{ER:85}.

Although the definition (\ref{eqn:lcemax-def}) of $\mathsf{L}_{\text{max}}$
is conceptually transparent, it does not lend itself to direct numerical\
implementation since in chaotic systems any initial deviation $\lVert \delta 
{\mathbf{x}}_{0}\rVert $, no matter how small, will eventually evolve into a
number $\lVert \delta {\mathbf{x}}_{t}\rVert $ which far exceeds the
(floating point) representation capabilities of computers. Hence more
indirect numerical approaches are called for, and several different ones
have been developed over the last decades \cite{JF:83;WSSV:85;SS:85;GK:87}.
In the following, we will adopt the Jacobian method which integrates the
equation for the time evolution of the deviation between two initially
neighboring orbits, linearizing it anew at each time along the orbits and
therefore having to manipulate only the relatively small deviations produced
by one time step. A generalization of this method to the calculation of the
whole non-negative Lyapunov spectrum was developed in Refs.~\cite{BGGS:80,SN:79}.
Our discussion and implementation follows Ref.~\cite{MH:03}.

The Jacobian method is most transparently formulated by rewriting our system
(\ref{eqn:movf1})--(\ref{eqn:movf5}) of four second-order differential
equations into a system of eight autonomous first-order equations 
\begin{equation}
\dot{{\mathbf{x}}}\!\left( t\right) ={\mathbf{F}}\!\left( {\mathbf{x}}%
\!\left( t\right) \right) ,  \label{eqn:ds}
\end{equation}%
where ${\mathbf{x}}$ comprises the four coordinates and their time
derivatives, i.e. ${\mathbf{x}}$ and ${\mathbf{F}}$ are eight-dimensional
column vectors. Under the initial conditions ${\mathbf{x}}\!\left(
t_{0}=0\right) ={\mathbf{x}}_{0}$, the unique solution of Eq.~\eqref{eqn:ds}
is the orbit ${\mathbf{x}}\!\left( t\right) ={\mathbf{\Phi }}\!\left( t,{%
\mathbf{x}}_{0}\right) $ with ${\mathbf{x}}_{0}={\mathbf{\Phi }}\!\left( 0,{%
\mathbf{x}}_{0}\right) $. We now linearize the above system in the small
deviation $\delta {\mathbf{x}}\!\left( t\right) $ at any point ${\mathbf{x}}%
\!\left( t\right) $ of the trajectory by substituting ${\mathbf{x}}\!\left(
t\right) ={\mathbf{x}}\!\left( t\right) +\delta {\mathbf{x}}\!\left(
t\right) $ into Eq.~\eqref{eqn:ds} and neglecting terms of second-order in $%
\delta {\mathbf{x}}\!\left( t\right) $. The result is a linear system of
first-order equations 
\begin{equation}
\delta {\dot{\mathbf{x}}}\!\left( t\right) ={\mathbf{J}}\!\left( {\mathbf{x}}%
\!\left( t\right) \right) \!\cdot \mspace{1.mu}\!\delta {\mathbf{x}}\!\left(
t\right)  \label{eqn:dtev}
\end{equation}%
for the deviation $\delta {\mathbf{x}}$ which contains the Jacobi matrix ${%
\mathbf{J}}\!\left( {\mathbf{x}}\!\left( t\right) \right) =\left[ \partial {{%
\mathbf{F}}\!\left( {\mathbf{x}}\right) }/\partial {{\mathbf{x}}}\right]
\!|_{{\mathbf{x}}={\mathbf{x}\left( t\right) }}$. In terms of the tangent
vector ${\boldsymbol{\xi}}\!\left( t\right) =\lim_{\lVert \delta {\mathbf{x}}%
\left( 0\right) \rVert \rightarrow 0}\delta {\mathbf{x}}\!\left( t\right)
\!/\lVert \delta {\mathbf{x}}\!\left( 0\right) \mspace{1.82mu}\!\rVert $ on
the orbit at ${\mathbf{x}}\!\left( t\right) $, Eq.~(\ref{eqn:dtev}) turns
into the so-called variational system for ${\boldsymbol{\xi}}\!\left(
t\right) $ under the initial data ${\mathbf{x}}_{0}$, 
\begin{equation}
{\dot{\boldsymbol{\xi}}}\!\left( t\right) ={\mathbf{J}}\!\left( {\mathbf{%
\Phi }}\!\left( t,{\mathbf{x}}_{0}\right) \right) \!\cdot \mspace{1.mu}\!{%
\boldsymbol{\xi}}\!\left( t\right) .  \label{eqn:veq}
\end{equation}%
It propagates small variations tangent to the orbit at time $t_{0}$ to small
variations tangent to the orbit at time $t$. In terms of the ${{%
\boldsymbol{\xi}}}${{, }Eq.}${~\eqref{eqn:lcemax-def}}$ {for }${\mathsf{L}_{%
\text{max}}}$ becomes 
\begin{equation}
\mathsf{L}_{\text{max}}\!\left( {\mathbf{x}}_{0},{\boldsymbol{\xi}}%
_{0}\right) =\lim_{t\rightarrow \infty }\chi \!\left( t\right) \text{ \ \ \
where \ \ \ \ }\chi \!\left( t\right) :=\frac{\ln {\lVert {\boldsymbol{\xi}}%
\!\left( t\right) \mspace{1.82mu}\!\rVert }}{t}.  \label{eqref}
\end{equation}%
Hence, one can calculate $\mathsf{L}_{\text{max}}$ by numerically
integrating the variational Eqs.~(\ref{eqn:veq}) up to sufficiently
large times. In order to supply the necessary input, i.e. the Jacobian ${%
\mathbf{J}}\!\left( {\mathbf{\Phi }}\!\left( t,{\mathbf{x}}_{0}\right)
\right) $ at each time step, one has to solve the equations of motion (\ref%
{eqn:ds}) for the orbit ${\mathbf{x}}\!\left( t\right) ={\mathbf{\Phi }}%
\!\left( t,{\mathbf{x}}_{0}\right) $ in parallel. A biased choice for the
initial tangent vector ${\boldsymbol{\xi}}_{0}$ is avoided by selecting its
orientation randomly. The long-term time evolution of the orbits and ${{%
\boldsymbol{\xi}}\!}\left( t\right) $ will then be dominated by the largest
Lyapunov exponent \footnote{%
For a more detailed understanding of chaotic systems, the remaining
(nonmaximal) Lyapunov exponents are also of interest. Their numerical
evaluation is significantly more involved, however \cite{BGGS:80,SN:79,MH:03}.}.

The numerical evaluation of Eq.~(\ref{eqref}) requires some additional
precaution, however, since for chaotic orbits the norm ${\lVert {%
\boldsymbol{\xi}}\!}\left( t\right) {\mspace{1.82mu}\!\!\rVert }$ will grow
large enough to generate floating point overflows on computers. To
circumvent these, we directly calculate the value of the required
logarithmic norm after $n$ time steps ($t_{n}=n\Delta t$) as the sum 
\begin{equation}
\ln {\lVert {\boldsymbol{\xi}}\!}\left( t_{n}\right) {\mspace{1.82mu}%
\!\!\rVert }=\sum_{i=1}^{n}\ln \frac{{\lVert {\boldsymbol{\xi}}\!}\left(
t_{i}\right) {\mspace{1.82mu}\!\!\rVert }}{{\lVert {\boldsymbol{\xi}}\!}\left(
t_{i-1}\right) {\mspace{1.82mu}\!\!\rVert }}
\end{equation}%
(with ${\lVert {\boldsymbol{\xi}}\!}\left( t_{0}\right) {\mspace{1.82mu}%
\!\!\rVert }=\lVert {\boldsymbol{\xi}}_{0}\mspace{1.82mu}\!\rVert =1$) of
logarithmic length increments after each time step \cite{ras90}. The
incremental norm changes ${\lVert {\boldsymbol{\xi}}\!}\left( t_{i}\right) {%
\mspace{1.82mu}\!\!\rVert /\lVert {\boldsymbol{\xi}}\!}\left( t_{i-1}\right) {%
\mspace{1.82mu}\!\!\rVert }$ remain small enough to be representable by
floating point numbers and are obtained by renormalizing $\boldsymbol{\xi}$
to unit length after each evaluation. (Of course, the renormalized $%
\boldsymbol{\xi}$ remain solutions of the linear variational system (\ref%
{eqn:veq}).)

\begin{figure}[tbp]
\begin{center}
\includegraphics[height = 9cm]{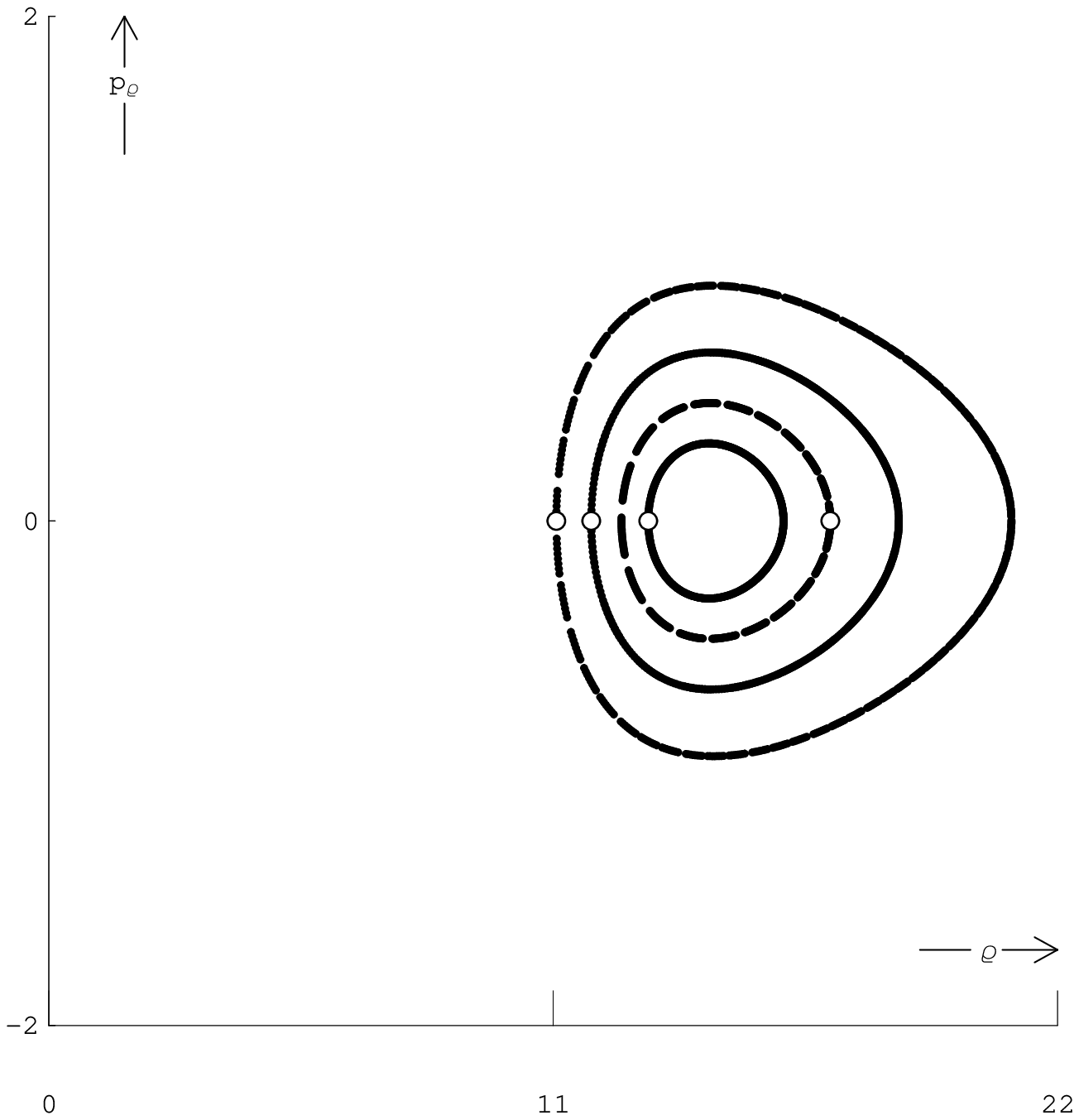}
\end{center}
\caption{Poincar\'{e} sections, in the $(\protect\varrho,p_{\protect\varrho%
}) $ plane located at the position $\mathcal{M}_{1} = 0$, of four orbits
from the Hamiltonian flow of Eq.~\eqref{h} with $\mathcal{M}^2=1906.71$.}
\label{fig:pp01}
\end{figure}

\begin{figure}[tbp]
\begin{center}
\includegraphics[height = 9cm]{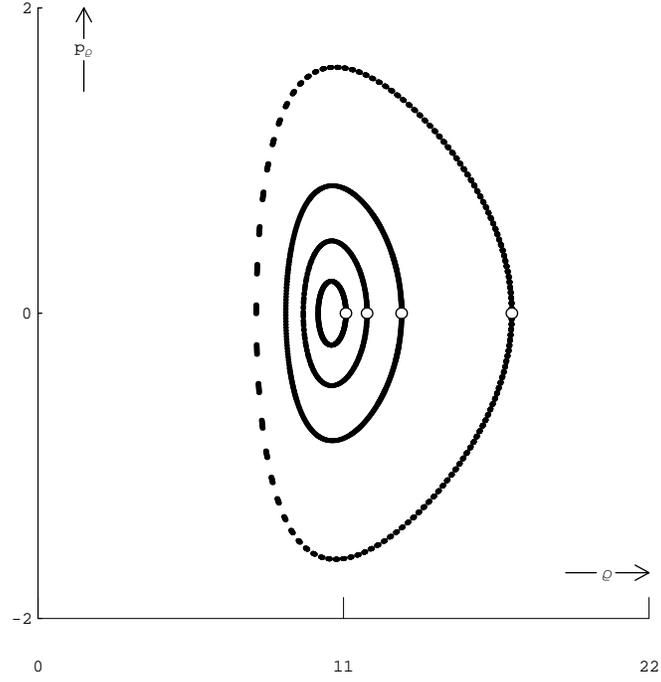}
\end{center}
\caption{Same as Fig.~\ref{fig:pp01} with $\mathcal{M}^2=2152.95$.}
\label{fig:pp02}
\end{figure}

\begin{figure}[tbp]
\begin{center}
\includegraphics[height = 9cm]{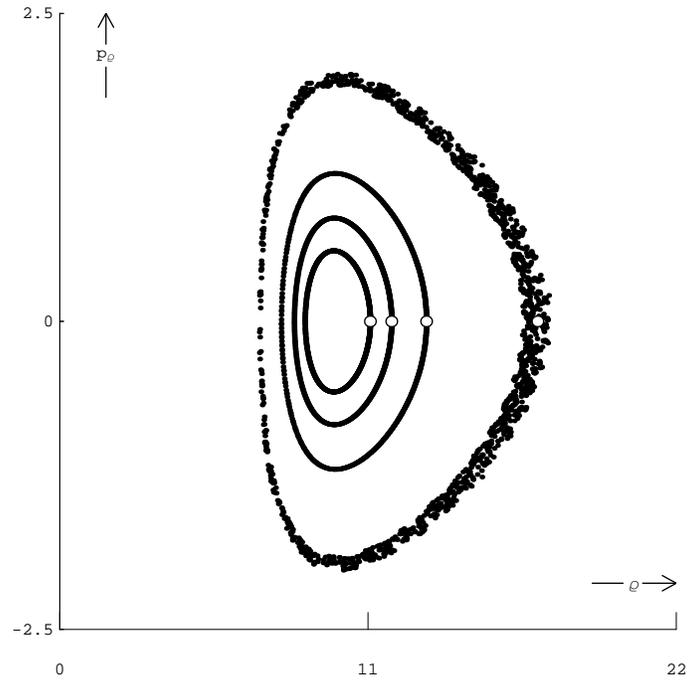}
\end{center}
\caption{Same as Fig.~\ref{fig:pp01} with $\mathcal{M}^2=2237.7$. (The outermost section
contains about 13500 points.)}
\label{fig:pp03}
\end{figure}

\begin{figure}[tbp]
\begin{center}
\includegraphics[width = 18.cm]{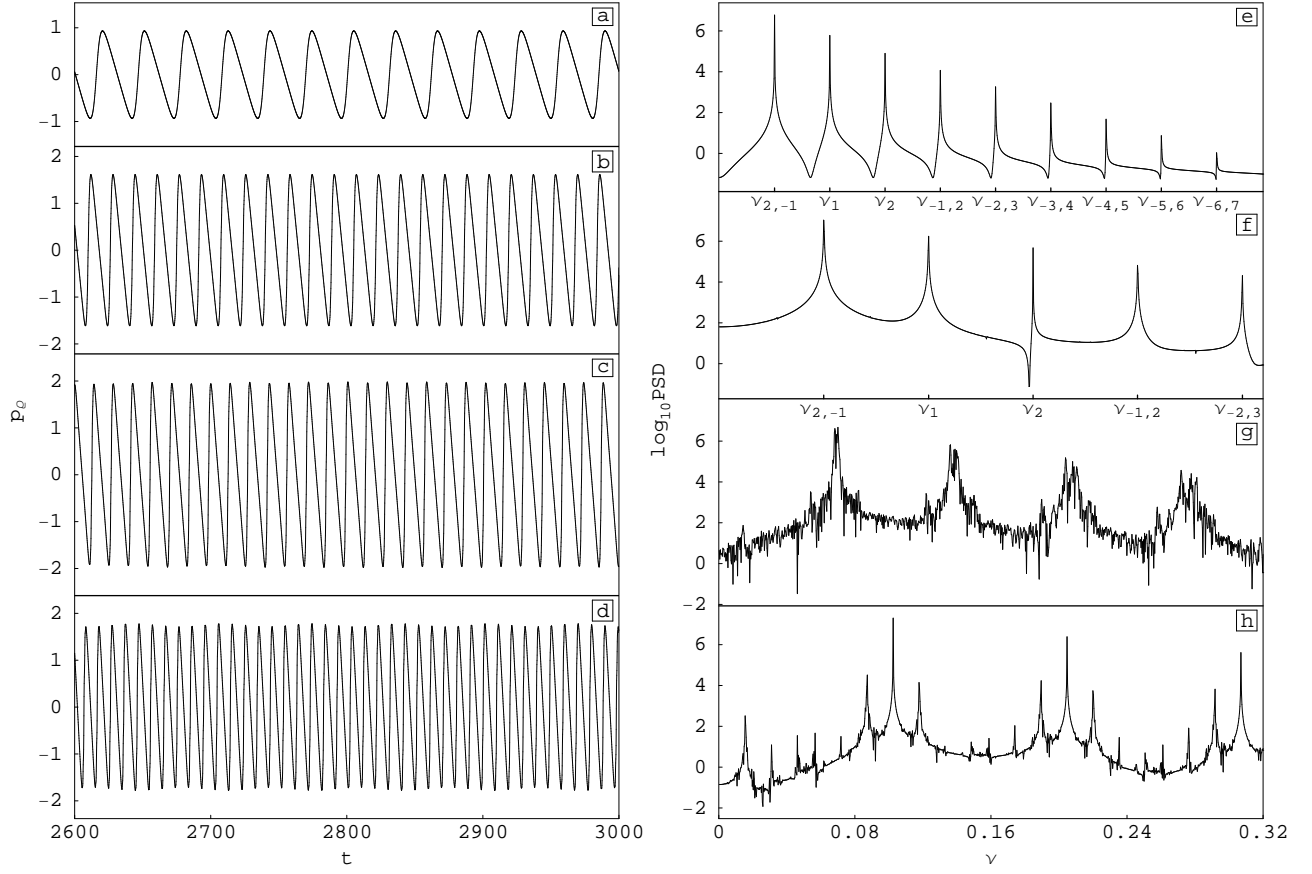}
\end{center}
\caption{Panel a (b, c) shows the time dependence of $p_\protect\protect%
\varrho$ and panel e (f, g) the logarithm of the corresponding power spectrum for the
dyon-pair orbit whose Poincar\'e section is plotted in Fig.~\ref{fig:pp01} (%
\ref{fig:pp02}, \ref{fig:pp03}). The panels d and h show $p_\protect%
\protect\varrho(t)$ and the corresponding logarithmic power spectrum for the additional
chaotic orbit described in the text.}
\label{fig:qtps}
\end{figure}

We now calculate, by means of the Jacobian technique, the
maximal Lyapunov exponents of the four dyon-pair orbits whose power spectra
were obtained in the last section and plotted in Figs.~\ref{fig:qtps} e-h.
The Poincar\'{e} sections of the first three of these orbits are the
outermost in Figs.~\ref{fig:pp01}--\ref{fig:pp03}. The integration range
consists of $n=2\times 10^{8}$ time steps which corresponds to approximately 
$650$ periods for the outermost quasiperiodic orbit of Fig.~\ref{fig:pp01}.
In order to monitor the time evolution of $\chi $ during its approach to the
limiting value $\mathsf{L}_{\text{max}}$ according to Eq.~(\ref{eqref}), we
plot $\chi \!\left( t\right) $ for each of the four orbits in Fig.~\ref%
{chlyp}. Obviously, there is a striking qualitative difference between the
plots for the two orbits which we had identified as (two-frequency)
quasiperiodic in Sec.~\ref{psp} (with power spectra in Figs.~\ref{fig:qtps}e
and~\ref{fig:qtps}f) and the irregular ones whose power spectra are shown in
Figs.~\ref{fig:qtps}g and~\ref{fig:qtps}h. For the quasiperiodic orbits one
infers within numerical uncertainties that $\mathsf{L}_{\text{max}%
}=\lim_{\,t\rightarrow \infty }\mspace{1.82mu}\!\chi \!\left( t\right) =0$
while $\chi \!\left( t\right) $ approaches the finite and positive values 
\begin{equation}
\mathsf{L}_{\text{max,1}}\sim 0.02,\text{ \ \ \ \ \ \ }\mathsf{L}_{\text{%
max,2}}\sim 0.008
\end{equation}%
for the two aperiodic orbits. Around these orbits the motion is therefore
exponentially sensitive to small variations of the initial conditions in at
least one phase-space direction. This is the prototypical hallmark of chaos.
We have thus achieved two of our main objectives, namely, the unequivocal
confirmation\ of the chaoticity of the two-dyon system and a (semi-)
quantitative determination of its primary characteristic scales.

The maximal Lyapunov exponent $\mathsf{L}_{\text{max,1}}$ of the first
chaotic orbit (with the power spectrum in Fig.~\ref{fig:qtps}g) is more than
twice as large as that of the second chaotic orbit (with the power spectrum
in Fig.~\ref{fig:qtps}h), $\mathsf{L}_{\text{max,2}}$, although the
differences in the initial conditions ($\mathcal{M}^{2}=2237.70$ and $\beta
=3.14$ for the first and $\mathcal{M}^{2}=2359.46$ and $\beta =3.13$ for the
second orbit) seem comparatively small. Closer inspection of these two
orbits reveals that the minimal and maximal dyon distances in the first one
are $\varrho _{\min ,1}=6.99$ and $\varrho _{\max ,1}=17.98$ while the
corresponding momenta $p_{\varrho }$ vary inbetween $p_{\varrho ,1\min /\max
}=\pm 2.09$. For the second chaotic orbit one has a smaller
variation between minimal and maximal $\varrho $ values, $\varrho _{\min
,2}=6.73$ and $\varrho _{\max ,2}=14.23$, and a somewhat smaller range of
momenta, bounded by $p_{\varrho ,2\min /\max }=\pm 1.94$. Although the
initial Coulomb attraction between the dyons is stronger in the second orbit
(since its total angular momentum squared $\mathcal{M}^{2}$ is 5\% larger)
and the dyons therefore come closer to each other and depart farther from
the integrable asymptotic domain, its chaoticity---as measured by the
maximal Lyapunov exponent---is still considerably smaller. This might be a
consequence of the generally smaller momenta of the second orbit.

\begin{figure}[tbp]
\begin{center}
\includegraphics[width = 14.cm]{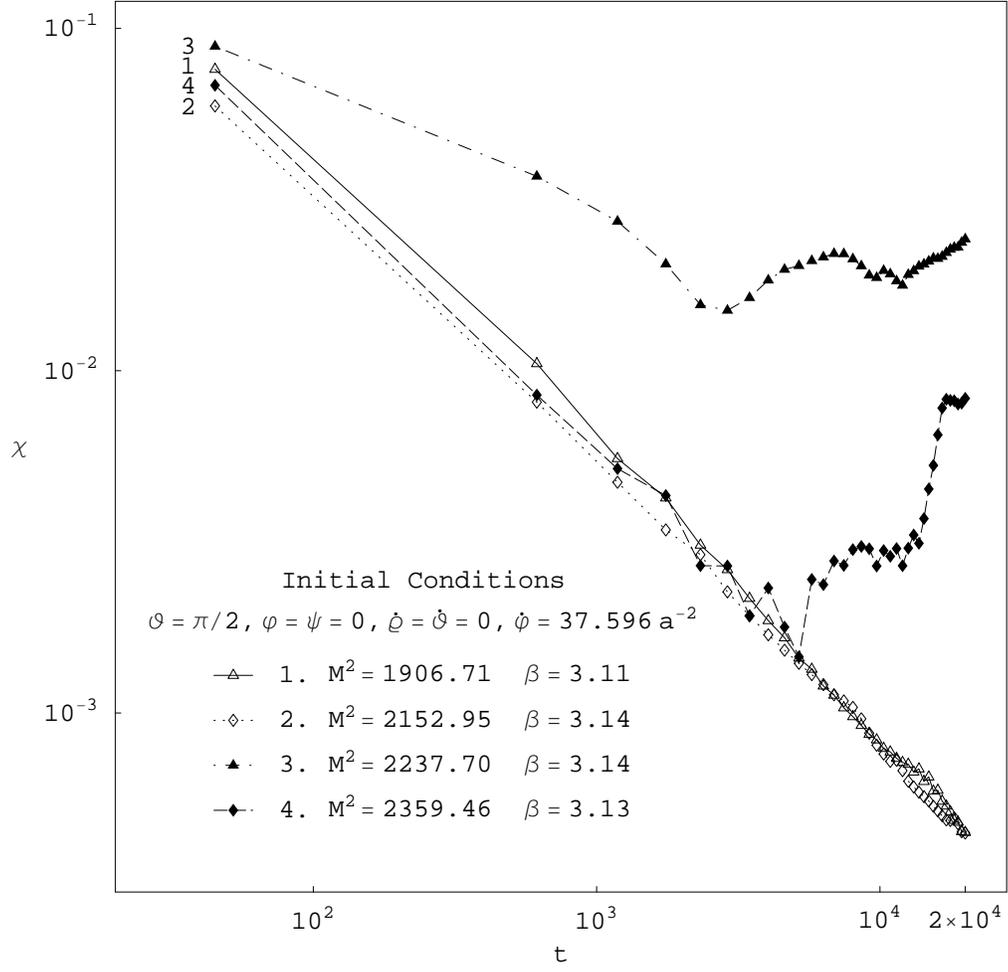}
\end{center}
\caption{The full (dotted, dash-dotted, dashed) curve depicts $\protect\chi%
(t)$ for the dyon-pair orbit whose power spectrum is plotted in Fig.~\ref%
{fig:qtps} e (f, g, h).}
\label{chlyp}
\end{figure}

The behavior of $\chi $ also reveals a conspicuous qualitative difference
between the two chaotic orbits. While the $\chi \!\left( t\right) $ of the
first one (dot-dashed curve with filled triangles in Fig.~\ref{chlyp})
clearly stays above those of the quasiperiodic orbits for all $t$, the $\chi
\!\left( t\right) $ of the second one (dashed curve with filled squares in
Fig.~\ref{chlyp}) follows its quasiperiodic counterparts for a long time
rather closely and then suddenly rises in a ``burstlike'' onset of chaos.
This intriguing behavior may be a first vestige of intermittency in the
geodesic dyon-pair motion. It also seems to explain the more regular and
approximately quasiperiodic power spectrum of the second orbit [Fig.~\ref%
{fig:qtps}h]. Stronger evidence for the potentially intermittent behavior
could be established by identifying complete intermittency intervals of $%
\chi \!\left( t\right) $ but would require much longer orbits. The power
spectra of Sec.~\ref{psp} provide further testing grounds for
intermittency which could be exploited, e.g., by subjecting them to a moment analysis or by searching
for a power-law behavior in their low-frequency tails (i.e. $1/f$ noise) %
\cite{ben85}.

\begin{figure}[tbp]
\begin{center}
\includegraphics[width = 14.cm]{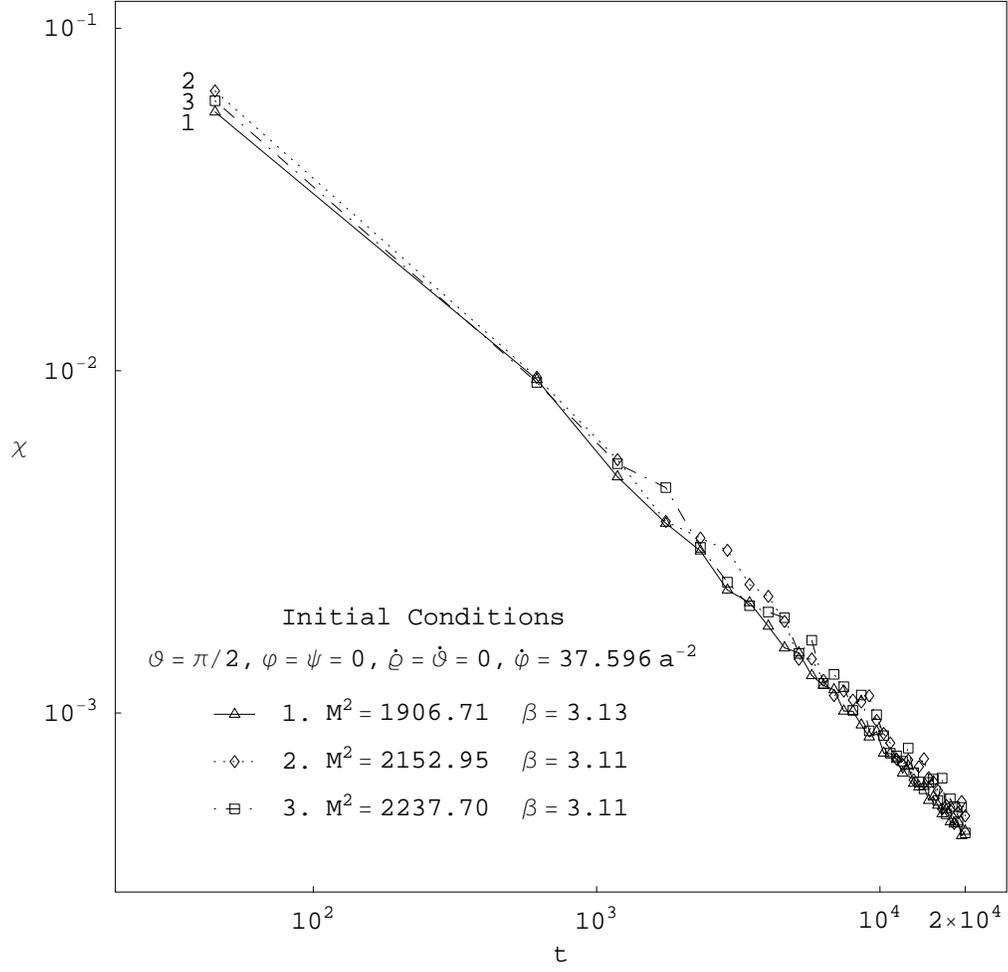}
\end{center}
\caption{The function $\protect\chi(t)$ for three quasiperiodic dyon-pair
orbits. The full (dotted, dash-dotted) curve corresponds to the orbit with
the innermost Poincar\'{e} section in Fig.~\ref{fig:pp01} (\ref{fig:pp02}%
, \ref{fig:pp03}).}
\label{reglyp}
\end{figure}

Although the large-time limit of $\chi \!\left( t\right) $ (i.e. the maximal
Lyapunov exponent)\ vanishes for quasiperiodic orbits, its characteristic
time dependence may contain useful information as well. In order to examine
the behavior of $\chi \!\left( t\right) $ for quasiperiodic motion more
closely, we select a new sample of three quasiperiodic orbits, namely, those
whose Poincar\'{e} sections are the \emph{inner}most in each of the Figs.~%
\ref{fig:pp01}--\ref{fig:pp03}, and display the corresponding $\chi \!\left(
t\right) $ in Fig.~\ref{reglyp}. (The full (dotted, dash-dotted)\ line
corresponds to the orbit with the innermost Poincar\'{e} section in Fig.~\ref%
{fig:pp01} (\ref{fig:pp02}, \ref{fig:pp03}).) At sufficiently large $t$ all
curves seem to approach straight lines, which indicates a power-law behavior 
\begin{equation}
\chi \!\left( t\right) \sim at^{-b}\text{ \ \ \ \ \ \ (}a,b>0\text{).}
\end{equation}%
This type of scaling behavior is frequently encountered in quasiperiodic
systems and supports the visual impression that all curves have indeed the
expected $\mathsf{L}_{\text{max}}=\lim_{\,t\rightarrow \infty }%
\mspace{1.82mu}\!\chi \!\left( t\right) =0$.

We close this section by recalling that (positive)\ maximal Lyapunov
exponents contain crucial information about the physical behavior even of
quantum systems. A notable example is their partial characterization of
nonequilibrium processes in semiclassical systems (e.g. at high
temperature) where they typically set the scale of relaxation times and
thermalization rates \cite{hei97}. One might expect that our Lyapunov
exponents play a similar role in determining, e.g., the equilibration rate
of a nonequilibrium dyon system.

\section{Summary and conclusions}

We have analyzed several representative motion patterns of two interacting
BPS dyons in the geodesic approxi\-ma\-tion. The main emphasis was put on
discerning regular and chaotic orbits and on characterizing
them both qualitatively and quantitatively by means of suitable chaos
indicators.

Our study is based on a sample of thirteen long-time phase space
trajectories for which four-dimensional time series were generated by
numerically integrating the equations of motion with high accuracy over
typically $2^{25}$ time steps. The initial data sets were chosen to cover a
representative range of motion patterns and to explore the low-energy dyon
interactions at different strengths. Hence the orbit set includes sequences
of trajectories whose decreasing minimal dyon separations interpolate
between asymptotic dyon distances, where charge exchange becomes ineffective
and the geodesic dynamics integrable, and relatively small minimal
separations for which the interactions are expected to become nonintegrable.

A second motive for our initial-data selection was to include orbits in the
vicinity of those for which Poincar\'{e} sections were already available.
This allows for a direct comparison with previous results and served as a
useful benchmark and starting point for our work. We constructed Poincar\'{e}
sections in the radial coordinate-momentum plane for twelve orbits.
They qualitatively confirm the earlier results and extend them to
trajectories in neighboring phase space regions. Moreover, they contain
useful graphical information on the shape of the constant-energy surfaces and
on the qualitative behavior of the two-dyon system as a function
of the initial conditions. The dimensionality of the Poincar\'{e}
sections, in particular, provides clear indications for the underlying
trajectories to be either quasiperiodic or chaotic.

In order to complement the results of the Poincar\'{e} section
analysis, we have additionally calculated high-resolution
power spectra of selected orbits. The spectral analysis
of the momentum conjugate to the dyon
separation has provided particularly clean distinctions between quasiperiodic
and aperiodic orbits which strengthen and extend the interpretation of the
Poincar\'{e} sections. In addition, the power spectra produced the first
quantitative characterization of quasiperiodic dyon-pair orbits by
establishing the number of their fundamental modes (two), determining their
frequencies and yielding the strength distribution over the various
harmonics. The emergence of just the minimal, i.e. two-mode quasiperiodicity
is rather widespread among sufficiently strongly coupled, nonlinear
dynamical systems. The common expectation that nonlinear couplings between
more than two fundamental modes increasingly turn quasiperiodicity into
chaos might therefore apply to the two-dyon system as well and explain why
we have only found two-mode-quasiperiodic and chaotic trajectories.

In contrast to their almost complete characterization of quasiperiodic motion
patterns, power spectra do extract relatively little pertinent and
quantitative information from irregular orbits. Hence we have additionally
calculated the primary characteristic scales of chaotic motion, i.e. the
maximal Lyapunov exponents, for a suitable subset of orbits. As expected, the
Lyapunov exponents of orbits previously identified as quasiperiodic were
found to vanish. The two orbits with an irregular broadband power spectrum,
on the other hand, have finite and positive maximal Lyapunov exponents whose
values were approximately determined as $\mathsf{L}_{\text{max,1}}\sim 0.02$
and $\mathsf{L}_{\text{max,2}}\sim 0.008$. Those provide our most
unequivocal and quantitative evidence for the chaoticity of the dyon-dyon
interactions. The orbit with the smaller Lyapunov exponent shows in addition
signs of intermittent behavior.

Reassuringly, the results of all three employed analysis methods, i.e.
Poincar\'{e} sections, power spectra and maximal Lyapunov exponents, are
fully consistent with each other. Taken together, they provide convincing
evidence for and a quantitative description of both quasiperiodic and
chaotic regions in the low-energy phase space of two BPS dyons. Moreover, the
integrability of noninteracting\ dyon systems allows to trace the origin
of the chaotic behavior to the interactions between the dyons. These interactions may
therefore help to disorder monopole ensembles similar to those which are
expected to populate the vacuum of the strong interactions. In any case, our
results imply that no more than the three explicitly known integrals of the
motion are conserved by the geodesic forces between the dyons.

\begin{acknowledgments}
This work was partially supported by CAPES, CNPq and FAPESP of Brazil.
\end{acknowledgments}

\appendix

\section{Atiyah-Hitchin metric}

\label{ahm}

In this appendix we establish our notation and briefly summarize pertinent
features of the Atiyah-Hitchin (AH) metric and its Christoffel connection on
the internal moduli space $M_{2}^{\left( 0\right) }$ of the
Bogomol'nyi-Prasad-Sommerfield (BPS) monopole pair. Full details can be
found in the book \cite{AH:88}.

BPS magnetic monopoles are regular classical soliton solutions of the SU$%
\left( 2\right) $ Yang-Mills-Higgs equations in the limit of vanishing Higgs
self-coupling \cite{pra75,bog76}. The static two-monopole solutions define,
as explained in Sec.~\ref{clasmo}, the coollective-coordinate or moduli
manifold $M_{2}^{\left( 0\right) }$. Its metric was first written down
explicitly by Atiyah and Hitchin \cite{AH:85} and determines the geodesic
dynamics of two interacting BPS dyons at small velocities. The AH
construction exploits the facts that the two-monopole moduli space is a
hyper-K\"{a}hler (or Hamiltonian) manifold~\cite{AH:88,JG:94}, that it
admits SO($3$) as a group of isometries and that the orbits under the action
of SO($3$) are with one exception three-dimensional. One can then show that
the metric on $M_{2}^{\left( 0\right) }$ must be of the form 
\begin{equation}
ds^{2}=f^{2}d\varrho ^{2}+\left( a^{2}l_{\alpha }l_{\beta }+b^{2}m_{\alpha
}m_{\beta }+c^{2}n_{\alpha }n_{\beta }\right) dx^{\alpha }dx^{\beta } ,
\label{eqn:metric}
\end{equation}%
where $a$, $b$, $c$ and $f$ are functions of the ``radial'' variable $%
\varrho $ only and $l_{\alpha }dx^{\alpha }$, $m_{\alpha }dx^{\alpha }$ and $%
n_{\alpha }dx^{\alpha }$ are differential forms on the three-sphere $S^{3}$
which can be taken as 
\begin{align}
& \sigma _{1}:=l_{\alpha }dx^{\alpha }=-\sin {\psi }\,d\vartheta +\sin {%
\vartheta }\cos {\psi }\,d\varphi ,  \notag \\
& \sigma _{2}:=m_{\alpha }dx^{\alpha }=\cos {\psi }\,d\vartheta +\sin {%
\vartheta }\sin {\psi }\,d\varphi ,  \notag \\
& \sigma _{3}:=n_{\alpha }dx^{\alpha }=d\psi +\cos {\vartheta }\,d\varphi ,
\label{eqn:ss}
\end{align}%
in terms of three Euler angles $\vartheta$, $\varphi$ and $\psi$ in the
intervals $0\leq \vartheta \leq \pi$, $0\leq \varphi \leq 2\pi$, $0\leq \psi
\leq 2\pi $. Accordingly, the two-monopole moduli space is parameterized by
a coordinate $\varrho $ which describes the separation between the
monopoles, two angular coordinates $\vartheta $ and $\varphi $ which
determine the orientation of the axis that joins the monopoles, and the
angle $\psi $ which fixes the position of the (generally axially asymmetric)
two-monopole system with respect to rotations around this axis.

By casting the line element~\eqref{eqn:metric} into the Riemannian form $%
g_{\alpha \beta }\,dx^{\alpha }dx^{\beta }$, one can straightforwardly
verify that the fundamental metric tensor $g_{\alpha \beta }$ is symmetric,
i.e., $g_{\alpha \beta }=g_{\beta \alpha }$, and that the contravariant
tensor $g^{\alpha \beta }$ is its inverse, $g_{\alpha \beta }\,g^{\beta
\gamma }=\delta _{\alpha }{}^{\gamma }$, as it should be. Explicitly, the
diagonal elements of $g_{\alpha \beta }$ read 
\begin{align}
g_{00}& =f^{2},\text{ \ \ \ \ }g_{33}=c^{2},  \notag \\
g_{11}& =a^{2}\sin ^{2}{\!\psi }+b^{2}\cos ^{2}{\!\psi },  \notag \\
g_{22}& =a^{2}\sin ^{2}{\!\vartheta }\cos ^{2}{\!\psi }+b^{2}\sin ^{2}{%
\!\vartheta }\sin ^{2}{\!\psi }+c^{2}\cos ^{2}{\!\vartheta } , \label{metdiag}
\end{align}%
while the nondiagonal ones are 
\begin{align}
g_{01}& =g_{02}=g_{03}=g_{13}=0,  \notag \\
g_{12}& =\left( b^{2}-a^{2}\right) \sin {\vartheta }\sin {\psi }\cos {\psi },
\notag \\
g_{23}& =c^{2}\cos {\vartheta }.  \label{eqn:mc}
\end{align}%
Similarly, for the elements of the contravariant tensor $g^{\alpha \beta }$
one has 
\begin{align}
g^{00}& =\frac{1}{f^{2}},\text{ \ \ \ \ }g^{33}=\frac{1}{c^{2}}+\left( \frac{%
\cos ^{2}{\!\psi }}{a^{2}}+\frac{\sin ^{2}{\!\psi }}{b^{2}}\right) \cot ^{2}{%
\!\vartheta },  \notag \\
g^{11}& =\frac{\cos ^{2}{\!\psi }}{b^{2}}+\frac{\sin ^{2}{\!\psi }}{a^{2}}, 
\notag \\
g^{22}& =\left( \frac{\cos ^{2}{\!\psi }}{a^{2}}+\frac{\sin ^{2}{\!\psi }}{%
b^{2}}\right) \csc ^{2}{\!\vartheta } , \label{metcovdiag}
\end{align}%
and%
\begin{align}
g^{01}& =g^{02}=g^{03}=0,  \notag \\
g^{12}& =\frac{\left( a^{2}-b^{2}\right) \csc {\vartheta }\sin {\psi }\cos {%
\psi }}{a^{2}b^{2}},  \notag \\
g^{13}& =\frac{\left( b^{2}-a^{2}\right) \cot {\vartheta }\sin {\psi }\cos {%
\psi }}{a^{2}b^{2}},  \notag \\
g^{23}& =-\left( \frac{\cos ^{2}{\!\psi }}{a^{2}}+\frac{\sin ^{2}{\!\psi }}{%
b^{2}}\right) \csc {\vartheta }\cot {\vartheta }.  \label{eqn:mc2}
\end{align}

With these expressions at hand, one can obtain explicit relations for the
functions $a,$ $b,$ $c$ and $f$ by means of the equation 
\begin{equation}
R_{\lambda \sigma }=\frac{\partial \Gamma ^{\alpha }{}_{\lambda \sigma }}{%
\partial x^{\alpha }}-\frac{\partial \Gamma ^{\alpha }{}_{\lambda \alpha }}{%
\partial x^{\sigma }}+\Gamma ^{\alpha }{}_{\nu \alpha }\Gamma ^{\nu
}{}_{\lambda \sigma }-\Gamma ^{\alpha }{}_{\nu \sigma }\Gamma ^{\nu
}{}_{\lambda \alpha }=0  \label{eqn:feqgf}
\end{equation}%
for the symmetric second-rank tensor $R_{\lambda \sigma }$ which represents
the fact that $M_{2}^{\left( 0\right) }$ is hyper-K\"{a}hler. The $\Gamma
^{\alpha }{}_{\lambda \sigma }$ are the components of the Christoffel
connection, i.e. the Christoffel symbols of the second kind, which are
related to the metric tensor by
\begin{equation}
\Gamma ^{\alpha }{}_{\lambda \sigma }=\frac{1}{2}\,g^{\alpha \mu }\left( 
\frac{\partial g_{\sigma \mu }}{\partial x^{\lambda }}+\frac{\partial
g_{\lambda \mu }}{\partial x^{\sigma }}-\frac{\partial g_{\lambda \sigma }}{%
\partial x^{\mu }}\right) .  \label{eqn:csymb}
\end{equation}%
(Obviously, they are symmetric under exchange of the lower indices.) After
inserting the expressions (\ref{metdiag})--(\ref{eqn:mc2}) for the elements
of the metric into Eq.~(\ref{eqn:feqgf}), one obtains the components of $%
R_{\lambda \sigma }$ as functions of $a,$ $b,$ $c$ and $f$: 
\begin{equation}
R_{00}=\left( \frac{a^{\prime }}{a}+\frac{b^{\prime }}{b}+\frac{c^{\prime }}{%
c}\right) \frac{f^{\prime }}{f}-\frac{a^{\prime \prime }}{a}-\frac{b^{\prime
\prime }}{b}-\frac{c^{\prime \prime }}{c}  \label{eqn:r00}
\end{equation}%
(the prime denotes differentiation with respect to the coordinate $\varrho $%
) and 
\begin{align}
R_{11}& =\Pi _{1}\sin ^{2}{\!\psi }+\Pi _{2}\cos ^{2}{\!\psi },  \notag \\
R_{12}& =\left( \Pi _{2}-\Pi _{1}\right) \sin {\vartheta }\sin {\psi }\cos {%
\psi },  \notag \\
R_{22}& =\left( \Pi _{1}\cos ^{2}{\!\psi }+\Pi _{2}\sin ^{2}{\!\psi }\right)
\sin ^{2}{\!\vartheta }+\Pi _{3}\cos ^{2}{\!\vartheta },  \notag \\
R_{23}& =\Pi _{3}\cos {\vartheta },\hspace{0.5cm}R_{33}=\Pi _{3},
\end{align}%
where 
\begin{align}
\Pi _{1}& =\frac{a^{4}-\left( b^{2}-c^{2}\right) ^{2}}{2b^{2}c^{2}}+\left[
\left( \frac{f^{\prime }}{f}-\frac{b^{\prime }}{b}-\frac{c^{\prime }}{c}%
\right) a^{\prime }-a^{\prime \prime }\right] \frac{a}{f^{2}},  \notag \\
\Pi _{2}& =\frac{b^{4}-\left( a^{2}-c^{2}\right) ^{2}}{2a^{2}c^{2}}+\left[
\left( \frac{f^{\prime }}{f}-\frac{a^{\prime }}{a}-\frac{c^{\prime }}{c}%
\right) b^{\prime }-b^{\prime \prime }\right] \frac{b}{f^{2}},  \notag \\
\Pi _{3}& =\frac{c^{4}-\left( a^{2}-b^{2}\right) ^{2}}{2a^{2}b^{2}}+\left[
\left( \frac{f^{\prime }}{f}-\frac{a^{\prime }}{a}-\frac{b^{\prime }}{b}%
\right) c^{\prime }-c^{\prime \prime }\right] \frac{c}{f^{2}}.
\end{align}%
All other components of $R$ (with the exception of those which differ from
the above by exchanging the indices) vanish. The equations (\ref{eqn:feqgf})
therefore reduce to 
\begin{equation}
\Pi _{1}=0,\hspace{0.5cm}\Pi _{2}=0,\hspace{0.5cm}\Pi _{3}=0,  \label{eqn:pis}
\end{equation}%
together with 
\begin{equation}
\frac{a^{\prime }}{a}\frac{b^{\prime }}{b}+\frac{b^{\prime }}{b}\frac{%
c^{\prime }}{c}+\frac{c^{\prime }}{c}\frac{a^{\prime }}{a}=\frac{1}{2}\left( 
\frac{1}{a^{2}}+\frac{1}{b^{2}}+\frac{1}{c^{2}}-\frac{a^{4}+b^{4}+c^{4}}{%
2a^{2}b^{2}c^{2}}\right) f^{2}.  \label{eqn:const}
\end{equation}%
The last equation, however, is just a first integral of Eqs.~\eqref{eqn:pis}
and may be regarded as a constraint imposed on the initial values of $a$, $b$%
, $c$ and its derivatives. The second-order equations, which of course
conserve this constraint, can be obtained from each other by cyclic
permutation of $\left( a,b,c\right) $. All four equations~\eqref{eqn:pis}
and \eqref{eqn:const} together constitute the vacuum Einstein equations of
the AH metric.

A particular set of first integrals of the second-order equations (\ref%
{eqn:pis}) are the first-order equation 
\begin{equation}
\frac{2bc}{f}\frac{da}{d\varrho }=\left( b-c\right) ^{2}-a^{2}
\label{eqn:abc}
\end{equation}%
and the two others which are obtained from it by cyclical permutation of $%
\left( a,b,c\right) $. These differential equations for the functions $a$, $%
b $, $c$ and $f$ were first derived in Ref.~\cite{BGPP:78}. They have been
linearized and solved in terms of Legendre's complete elliptic integrals of
the first and second kind in Ref.~\cite{AH:85} for $f=abc$ and in Ref.~\cite%
{GM:86} for $f=-b/\varrho $. Below we will adopt the second choice, $%
f=-b/\varrho $, since it leads to expressions which are more convenient for
our purposes. The explicit solution to Eq.~\eqref{eqn:abc} and its
permutations is then \cite{GM:86} 
\begin{align}
a^{2}& =4K(K-E)(E-\Bar{q}^{2}K)/E,  \notag \\
b^{2}& =4KE(K-E)/(E-\Bar{q}^{2}K),  \notag \\
c^{2}& =4KE(E-\Bar{q}^{2}K)/(K-E),  \label{eqn:abc2}
\end{align}%
where 
\begin{equation}
K\!\left( q\right) =\int_{0}^{\pi /2}d\tau \,(1-q^{2}\sin ^{2}{\!\tau }%
)^{-1/2},\hspace{0.5cm}E\!\left( q\right) =\int_{0}^{\pi /2}d\tau
\,(1-q^{2}\sin ^{2}{\!\tau })^{1/2}  \label{ke}
\end{equation}%
are the complete elliptic integrals of the first and second kind, $q$ is
related to $\varrho $ by $\varrho =2K(q)$ for $\pi \leq \varrho <\infty $
and $\Bar{q}=\sqrt{1-q^{2}}$ is the conjugate modulus. At the value $\varrho
=\pi $ (which implies $q=0$) the metric has a coordinate singularity, the
so-called ``bolt'', since $a(\pi )=0$ implies that the line element (\ref%
{eqn:metric}) becomes independent of $\sigma _{1}$.

\section{Equations and integrals of motion}

\label{eiom}

\subsection{Lagrange equations of motion}

The geodesic low-energy dynamics of the two-dyon system is governed by the
Lagrangian 
\begin{equation}
L=\frac{m}{2}g_{\alpha \beta }\dot{x}^{\alpha }\dot{x}^{\beta }=\frac{1}{2}%
\left( f^{2}\dot{\varrho}^{2}+a^{2}\omega _{x}^{2}+b^{2}\omega
_{y}^{2}+c^{2}\omega _{z}^{2}\right) , \label{eqn:lag}
\end{equation}%
where the reduced mass $m$ of the dyon-pair has been set to $m=1$\ (instead
of $m=2\pi $ in Ref.~\cite{GM:86}). The first equation in (\ref{eqn:lag})
describes generic geodesic motion while the second one is specialized to the
Atiyah-Hitchin metric on the internal collective coordinate manifold $%
M_{2}^{\left( 0\right) }$ in terms of the functions $a$, $b$, $c$ and $f$.
The components $\omega _{x}$, $\omega _{y}$ and $\omega _{z}$ of the
instantaneous (or ``body-fixed'') angular velocity $\Vec{\omega}$ along the
axes $x$, $y$ and $z$ may be expressed in terms of the rates of change of
the Euler angles as 
\begin{equation}
\omega _{x}\equiv \sigma _{1}/dt,\hspace{0.5cm}\omega _{y}\equiv \sigma
_{2}/dt,\hspace{0.5cm}\omega _{z}\equiv \sigma _{3}/dt  \label{eqn:omegas}
\end{equation}%
(cf. Eq.~(\ref{eqn:ss})). The canonically conjugate momenta associated with
the four relative coordinates $\varrho $, $\vartheta $, $\varphi $ and $\psi 
$ are 
\begin{align}
p_{\varrho }& =f^{2}\dot{\varrho},\hspace{0.5cm}p_{\psi }=c^{2}\omega _{z}, 
\notag \\
p_{\vartheta }& =b^{2}\omega _{y}\cos {\psi }-a^{2}\omega _{x}\sin {\psi }, 
\notag \\
p_{\varphi }& =a^{2}\omega _{x}\sin {\vartheta }\cos {\psi }+b^{2}\omega
_{y}\sin {\vartheta }\sin {\psi }+c^{2}\omega _{z}\cos {\vartheta }.
\label{eqn:ps}
\end{align}

The Euler-Lagrange equations for the time evolution of the collective
coordinates are obtained by varying the action based on the Lagrangian (\ref%
{eqn:lag}). Variation with respect to the radial coordinate $\varrho \!\left(%
t\right) $ leads to%
\begin{equation}
f^{2}\ddot{\varrho}+ff^{\prime }\dot{\varrho}^{2}-a\,a^{\prime }\omega
_{x}^{2}-b\,b^{\prime }\omega _{y}^{2}-c\,c^{\prime }\omega _{z}^{2}=0,
\end{equation}%
while the equation of motion for the coordinate $\psi \!\left( t\right) $
becomes 
\begin{equation}
2\mspace{1.5mu}c\,c^{\prime }\omega _{z}\dot{\varrho}+c^{2}\dot{\omega}%
_{z}=\left( a^{-2}-b^{-2}\right) a^{2}b^{2}\omega _{x}\omega _{y}.
\label{eqn:mov3}
\end{equation}%
The remaining two equations of motion for $\varphi \!\left( t\right) $ and $%
\vartheta \!\left( t\right) $ are 
\begin{eqnarray}
\lefteqn{2\mspace{1.5mu}b\,b^{\prime }\omega _{y}\dot{\varrho}\cos \psi -2%
\mspace{1.5mu}a\,a^{\prime }\omega _{x}\dot{\varrho}\sin \psi +c^{2}\omega
_{z}\dot{\varphi}\sin \vartheta }  \notag \\
&=&\left( \dot{\omega}_{x}\sin {\psi }+\omega _{x}\omega _{z}\cos {\psi }%
\right) a^{2}-\left( \dot{\omega}_{y}\cos {\psi }-\omega _{y}\omega _{z}\sin 
{\psi }\right) b^{2}  \label{1}
\end{eqnarray}%
and 
\begin{eqnarray}
\lefteqn{2\mspace{1.5mu}a\,a^{\prime }\omega _{x}\dot{\varrho}\sin \vartheta
\cos \psi +2\mspace{1.5mu}b\,b^{\prime }\omega _{y}\dot{\varrho}\sin
\vartheta \sin \psi +2\mspace{1.5mu}c\,c^{\prime }\omega _{z}\dot{\varrho}%
\cos \vartheta }  \notag \\
&=&\bigl[\bigl(\dot{\psi}\sin {\vartheta }\sin {\psi }-\dot{\vartheta}\cos {%
\vartheta }\cos {\psi }\bigr)\mspace{1.5mu}\omega _{x}-\dot{\omega}_{x}\sin {%
\vartheta }\cos {\psi }\bigr]a^{2}  \notag \\
&&-\mspace{4.75mu}\bigl[\bigl(\dot{\psi}\sin {\vartheta }\cos {\psi }+\dot{%
\vartheta}\cos {\vartheta }\sin {\psi }\bigr)\mspace{1.5mu}\omega _{y}+\dot{%
\omega}_{y}\sin {\vartheta }\sin {\psi }\bigr]b^{2}  \notag \\
&&+\mspace{5.85mu}\bigl(\omega _{z}\dot{\vartheta}\sin {\vartheta }-\dot{%
\omega}_{z}\cos {\vartheta }\bigr)\mspace{1.25mu}c^{2}.  \label{2}
\end{eqnarray}%
A pair of equations which are algebraic consequences of Eqs.~(\ref{1}) and (%
\ref{2}), 
\begin{align}
2\mspace{1.5mu}b\,b^{\prime }\omega _{y}\dot{\varrho}+b^{2}\dot{\omega}_{y}&
=\left( c^{-2}-a^{-2}\right) a^{2}c^{2}\omega _{x}\omega _{z},
\label{eqn:mov2} \\
2\mspace{1.5mu}a\,a^{\prime }\omega _{x}\dot{\varrho}+a^{2}\dot{\omega}_{x}&
=\left( b^{-2}-c^{-2}\right) b^{2}c^{2}\omega _{y}\omega _{z},
\label{eqn:mov1}
\end{align}%
will be helpful in arriving at a more efficient formulation. Indeed, with $%
\dot{\varrho}=p_{\varrho }/f^{2}$ one also has 
\begin{equation}
\ddot{\varrho}=\frac{\dot{p}_{\varrho }f-2\mspace{1.5mu}p_{\varrho }\dot{%
\varrho}\mspace{1.5mu}f^{\prime }}{f^{3}}=\frac{\dot{p}_{\varrho }}{f^{2}}-%
\frac{2\mspace{1.5mu}p_{\varrho }^{2}\mspace{1.5mu}f^{\prime }}{f^{5}}
\end{equation}%
and, introducing the ($\varrho $-dependently)\ rescaled angular velocities
(in the body-fixed frame) 
\begin{equation}
\mathcal{M}_{1}=a^{2}\omega _{x},\hspace{0.75cm}\mathcal{M}_{2}=b^{2}\omega
_{y},\hspace{0.75cm}\mathcal{M}_{3}=c^{2}\omega _{z},
\end{equation}%
one can rewrite the four equations of (relative) motion concisely as 
\begin{align}
\dot{\mathcal{M}}_{1}& =\left( \frac{1}{b^{2}}-\frac{1}{c^{2}}\right) 
\mathcal{M}_{2}\,\mathcal{M}_{3},  \label{eqn:movf1} \\
\dot{\mathcal{M}}_{2}& =\left( \frac{1}{c^{2}}-\frac{1}{a^{2}}\right) 
\mathcal{M}_{3}\,\mathcal{M}_{1},  \label{eqn:movf2} \\
\dot{\mathcal{M}}_{3}& =\left( \frac{1}{a^{2}}-\frac{1}{b^{2}}\right) 
\mathcal{M}_{1}\,\mathcal{M}_{2},  \label{eqn:movf4} \\
\dot{p}_{\varrho }& =\frac{p_{\varrho }^{2}\mspace{1.5mu}f^{\prime }}{f^{3}}+%
\mathcal{M}_{1}^{2}\,\frac{a^{\prime }}{a^{3}}+\mathcal{M}_{2}^{2}\,\frac{%
b^{\prime }}{b^{3}}+\mathcal{M}_{3}^{2}\,\frac{c^{\prime }}{c^{3}}.
\label{eqn:movf5}
\end{align}%
In this form, the equations of motion were first obtained in Refs.~\cite%
{GM:86,BM:88}.

\subsection{Integrals of the motion\label{ioms}}

Three independent \footnote{%
Two integrals of the motion are independent if their mutual Poisson bracket
vanishes.} constants of the geodesic Atiyah-Hitchin motion are known
explicitly. The first one can be immediately identified by noting that $%
\varphi $ is a cyclic coordinate, i.e. that it does not appear explicitly in
the Lagrangian (\ref{eqn:lag}). The corresponding generalized momentum $%
p_{\varphi }$ is therefore an integral of the motion, i.e. $\dot{p}_{\varphi
}=0$.

The Hamiltonian $H$ is obtained by the standard Legendre transformation 
\begin{equation}
H=p_{\varrho }\dot{\varrho}+p_{\vartheta }\dot{\vartheta}+p_{\varphi }\dot{%
\varphi}+p_{\psi }\dot{\psi}-L
\end{equation}%
of the Lagrangian (\ref{eqn:lag}). Using Eqs.~(\ref{eqn:ps}) to express
velocities in terms of coordinates and their conjugate generalized momenta,
the Hamiltonian becomes 
\begin{eqnarray}
H &=&\frac{1}{2}\left[ \frac{p_{\varrho }^{2}}{f^{2}}+\frac{p_{\psi }^{2}}{%
c^{2}}+\left( \frac{\cos ^{2}{\!}\psi }{a^{2}}+\frac{\sin ^{2}{\!}\psi }{%
b^{2}}\right) \left( p_{\varphi }\csc \vartheta -p_{\psi }\cot \vartheta
\right) ^{2}\right]  \notag \\
&&+\frac{p_{\vartheta }}{2}\left[ \left( \frac{\sin ^{2}{\!}\psi }{a^{2}}+%
\frac{\cos ^{2}{\!}\psi }{b^{2}}\right) p_{\vartheta }+\frac{a^{2}-b^{2}}{%
a^{2}b^{2}}\left( p_{\varphi }\csc \vartheta -p_{\psi }\cot \vartheta
\right) \sin \left( 2\psi \right) \right] , \label{h}
\end{eqnarray}%
which assumes the same values as the Lagrangian since both are of purely
kinetic origin. Neither in $L$ nor, as a consequence, in $H$ does the time
coordinate $t$ appear explicitly. The corresponding Hamilton equation, $\dot{%
H}=\left\{ H,H\right\} =0$, therefore trivially implies that the total
relative energy $H$ of the two-dyon system is conserved and furnishes a
second integral of the motion.

The third constant of the motion is the square of the (both frame- and
body-fixed) total angular momentum, $\mathcal{M}^{2}$, which can be
expressed as the sum of the squares of the rescaled body-fixed angular
velocities, i.e. $\mathcal{M}^{2}=\mathcal{M}_{1}^{2}+\mathcal{M}_{2}^{2}+%
\mathcal{M}_{3}^{2}$. One may check that $\mathcal{M}^{2}$ is conserved by
adding up the equations obtained from multiplying Eq.~(\ref{eqn:movf1}) by $2%
\mathcal{M}_{1}$, Eq.~(\ref{eqn:movf2}) by $2\mathcal{M}_{2}$ and Eq.~(\ref%
{eqn:movf4}) by $2\mathcal{M}_{3}$. The result is 
\begin{equation}
2\mathcal{M}_{1}\dot{\mathcal{M}}_{1}+2\mathcal{M}_{2}\dot{\mathcal{M}}_{2}+2%
\mathcal{M}_{3}\dot{\mathcal{M}}_{3}=0
\end{equation}%
and therefore 
\begin{equation}
\frac{d}{dt}\mathcal{M}^{2}=0.
\end{equation}%
By using the explicit expressions%
\begin{equation}
\mathcal{M}_{1}=\frac{p_{\varphi }\cos {\psi }-p_{\psi
}\cos {\vartheta }\cos {\psi }-p_{\vartheta }\sin {\vartheta }\sin {\psi }}{%
\sin {\vartheta }},
\end{equation}%
\begin{equation}
\mathcal{M}_{2}=\frac{p_{\varphi }\sin {\psi }-p_{\psi
}\cos {\vartheta }\sin {\psi }+p_{\vartheta }\sin {\vartheta }\cos {\psi }}{%
\sin {\vartheta }},
\end{equation}%
and 
\begin{equation}
\mathcal{M}_{3}=p_{\psi }
\end{equation}%
for the rescaled angular velocities, one can re-express the total angular
momentum squared in terms of the Euler angles and their canonically
conjugate momenta as 
\begin{equation}
\mathcal{M}^{2}=\sum_{j=1}^{3}\mathcal{M}_{j}^{2}=p_{\vartheta
}^{2}-2p_{\varphi }p_{\psi }\cot {\vartheta }\csc {\vartheta }+\left(
p_{\varphi }^{2}+p_{\psi }^{2}\right) \csc ^{2}{\!\vartheta }.
\label{eqn:m2}
\end{equation}

For the investigation of chaos in the two-dyon system it is important to
note that a fourth integral of the motion appears in the $\varrho
\rightarrow \infty $ limit, i.e. if the dyons are far separated. Indeed, in
this case the Atiyah-Hitchin metric simplifies to the Euclidean Taub-NUT
metric \cite{GM:86} and the electric charge density, associated with the
canonical momentum $p_{\psi }=c^{2}\omega _{z}$, becomes a fourth constant
of the motion, i.e. $\dot{p}_{\psi }=0$.

The simple physical explanation of this result is that infinitely separated
dyons cannot exchange electric charge, so that in addition to their overall
charge also their individual charges become time-independent. In this case
there exist at least four independently conserved quantities \footnote{%
In fact, there are even more constants of the motion \cite{GM:86}. Those are
associated with the SO($4$)\ (SO($3,1$)) symmetry
of bounded (unbounded) geodesic motion in Euclidean Taub-NUT \cite{feh87},
which can be traced to the existence of a Killing-Yano tensor in this space %
\cite{gib87}.} in the eight-dimensional phase space, enough to render the
geodesic dynamics in Euclidean Taub-NUT space (Liouville-)\ integrable \cite%
{GM:86}. As a consequence, the asymptotic motion of two BPS dyons cannot be
chaotic. Moreover, the potential transition from regular motion at $\varrho
\rightarrow \infty $ to chaotic motion for finite, decreasing $\varrho $ may
be ``delayed'' according to KAM\ theory \cite{kam} by a stepwise dissolution
of the invariant tori.

\newpage

\newpage

\newpage

\newpage

\newpage

\end{document}